\documentclass{lmcs} %

\usepackage[utf8]{inputenc}
\usepackage{algorithm}
\usepackage[noend]{algpseudocode}
\usepackage{MnSymbol}
\usepackage{stmaryrd}
\usepackage{varwidth}
\usepackage{graphicx}

\usepackage{tikz}
\usetikzlibrary{automata, graphs,positioning,chains,arrows,decorations.pathmorphing}

\usepackage
[disable]
{todonotes}

\usepackage{subcaption}

\newcommand{\cristian}[1]{\todo[inline, color=red!20,,caption={}]{{\bf Cristian:} #1}}

\newcommand{\cO}{\mathcal{O}}

\newcommand{\bbN}{\mathbb{N}}

\renewcommand{\setminus}{\,\,\backslash\,\,}

\newcommand{\cA}{\mathcal{A}}

\newcommand{\longtrans}[1]{\overset{#1}{\longrightarrow}}
\newcommand{\sem}[1]{\llbracket#1\rrbracket}

\newcommand{\vars}{\operatorname{vars}}

\newcommand{\directaccessproblem}{\textsc{MSODirectAccess}[\prec]}
\newcommand{\dyndirectaccessproblem}{\textsc{DynamicMSODirectAccess}[\prec]}
\newcommand{\cnt}[1]{\#{#1}}
\newcommand{\range}[1]{\operatorname{range}(#1)}
\newcommand{\dom}[1]{\operatorname{dom}(#1)}

\newcommand{\SG}{\mathbb{G}} 
\newcommand{\word}{\operatorname{seq}}
\renewcommand{\prod}{\operatorname{prod}}
\newcommand{\size}{\operatorname{size}}
\newcommand{\lab}{\operatorname{label}}
\newcommand{\id}{\operatorname{id}}
\newcommand{\height}{\operatorname{height}}

\newcommand{\join}{\operatorname{join}}
\newcommand{\node}{\operatorname{node}}
\newcommand{\splita}{\operatorname{split}}

\newcommand{\dinit}{\operatorname{init}}
\newcommand{\dset}{\operatorname{set}}
\newcommand{\dout}{\operatorname{out}}

\newcommand{\sdb}{\operatorname{SDB}}
\newcommand{\sset}{\mathcal{S}}

\newcommand{\econcat}{\operatorname{concat}}
\newcommand{\esplit}{\operatorname{split}}
\newcommand{\ecut}{\operatorname{cut}}
\newcommand{\epaste}{\operatorname{paste}}

\renewcommand{\operatorname}[1]{\mathsf{#1}}

\algrenewcommand\algorithmicindent{1.3em}%
\algnewcommand\algorithmicforeach{\textbf{for each}}
\algdef{S}[FOR]{ForEach}[1]{\algorithmicforeach\ #1\ \algorithmicdo}

\NewDocumentCommand{\srun}{O{q_1} O{p_1} O{q_2} O{p_2}}{\ensuremath{\mathsf{\#runs}((#1,#2) \rightarrow (#3,#4))}}

\NewDocumentCommand{\br}{}{\ensuremath{\mathbf{r}}}
\NewDocumentCommand{\bX}{}{\ensuremath{\mathbf{X}}}
\NewDocumentCommand{\bY}{}{\ensuremath{\mathbf{Y}}}
\NewDocumentCommand{\semc}{O{\bX} O{\bY} O{\br} O{G}}{\ensuremath{\sem{#4}_{#1=#3,#2 \leq #3}}}
\NewDocumentCommand{\pathc}{O{u} O{v} O{\bX} O{\bY} O{\br} O{G}}{\ensuremath{\mathsf{Path}_{#3=#5,#4 \leq #5}(#1,#2)}}

\NewDocumentCommand{\dsword}{O{D}}{\ensuremath{#1.\mathsf{word}}}

\NewDocumentCommand{\dsinit}{O{w} O{\cA}}{\ensuremath{\mathsf{init}(#1,#2)}}
\NewDocumentCommand{\dsda}{O{i} O{D}}{\ensuremath{\mathsf{DA}(#2, #1)}}

\NewDocumentCommand{\dsconcat}{O{D_1} O{D_2}}{\ensuremath{\mathsf{concat}(#1, #2)}}
\NewDocumentCommand{\dssplit}{O{i} O{D}}{\ensuremath{\mathsf{split}(#2, #1)}}

\NewDocumentCommand{\dscut}{O{i} O{j} O{D}}{\ensuremath{\mathsf{cut}(#3, #1, #2)}}
\NewDocumentCommand{\dspaste}{O{i} O{D_1} O{D_2}}{\ensuremath{\mathsf{paste}(#2, #3, #1)}}

\begin{document}

\title[Dynamic direct access of MSO query evaluation over strings]{Dynamic direct access of \texorpdfstring{\\}{}MSO query evaluation over strings}

\author[P.~Bourhis]{Pierre Bourhis\lmcsorcid{0000-0001-5699-0320}}[a]	%

\author[F.~Capelli]{Florent Capelli\lmcsorcid{0000-0002-2842-8223}}[b]	%

\author[S.~Mengel]{Stefan Mengel\lmcsorcid{0000-0003-1386-8784}}[b]	%

\author[C.~Riveros]{Cristian Riveros\lmcsorcid{0000-0003-0832-116X}}[c]	%

\address{Univ. Lille, CNRS, Inria, Centrale Lille, UMR 9189 CRIStAL, F-59000 Lille, France}	%
\email{pierre.bourhis@univ-lille.fr}  %

\address{Univ. Artois, CNRS, UMR 8188, Centre de Recherche en Informatique de Lens (CRIL), F-62300 Lens, France}	%
\email{\{capelli, mengel\}@cril.fr}  %

\address{Pontificia Universidad Cat\'olica de Chile \& Millennium Institute for Foundational Research on Data }	%
\email{cristian.riveros@uc.cl}  %

	\maketitle

\begin{abstract}
  We study the problem of evaluating a Monadic Second Order (MSO) query over strings under updates in the setting of direct access. We present an algorithm that, given an MSO query with first-order free variables represented by an unambiguous variable-set automaton $\cA$ with state set $Q$ and variables $X$ and a string~$s$, computes a data structure in time $\cO(|Q|^\omega\cdot |X|^2 \cdot |s|)$ and, then, given an index $i$ retrieves, using the data structure, the $i$-th output of the evaluation of $\cA$ over $s$ in time $\cO(|Q|^\omega \cdot |X|^3 \cdot \log(|s|)^2)$ where $\omega$ is the exponent for matrix multiplication. Ours is the first efficient direct access algorithm for MSO query evaluation over strings; such algorithms so far had 
  only been studied for first-order queries and conjunctive queries over relational data.
    
  Our algorithm gives the answers in lexicographic order where, in contrast to the setting of conjunctive queries, the order between variables can be freely chosen by the user without degrading the runtime. Moreover, our data structure can be updated efficiently after changes to the input string, allowing more powerful updates than in the enumeration literature, e.g.~efficient deletion of substrings, concatenation and splitting of strings, and cut-and-paste operations.
  Our approach combines a matrix representation of MSO queries and a novel data structure for dynamic word problems over semi-groups which yields an overall algorithm that is elegant and easy to formulate. 

\end{abstract}

	\section{Introduction}\label{sec:introduction}

The aim of direct access algorithms in query answering is to represent query answers in a compact and efficiently computable way without materializing them explicitly, while still allowing efficient access as if the answers were stored in an array. This is modeled in a two-stage framework where, first, in a preprocessing phase, given an input, one computes a data structure which then, in the so-called access phase, allows efficiently accessing individual answers to the query by the position they would have in the materialized query result.

For this approach, there is a trade-off between the runtime of the preprocessing and the access phase: on the one hand, one could simply materialize the answers and then answer all queries in constant time; on the other hand, one could just not preprocess at all and then for every access request evaluate from scratch. When designing direct access algorithms, one thus tries to find interesting intermediate points between these extremes, mostly giving more importance to the access time than the preprocessing, since the latter has to be performed only once for every input while the former can be performed an arbitrary number of times.

Direct access algorithms were first introduced by Bagan, Durand, Grandjean, and Olive in~\cite{BaganDGO08} in the context of sampling and enumeration for first-order queries (both can easily be reduced to direct access). However, it was arguably the very influential thesis of Brault-Baron~\cite{BraultBaron13} that established direct access as an independent regime for query answering.
Since then, there has been a large amount of work extending and generalizing Brault-Baron's work in several directions~\cite{CarmeliZBCKS22,CarmeliTGKR23,BringmannCM22,EldarCK24,CapelliI24}.

Curiously absent from this line of work is the query evaluation of Monadic Second Order logic (MSO), which in other contexts, like enumeration algorithms, %
has often been studied in parallel with conjunctive queries. %
In this paper, we make a first step in that direction, giving a direct access algorithm for MSO queries with free first-order variables over strings.

Our approach to direct access decomposes into several steps: first, we reduce direct access to a counting problem by using binary search to find the answers to be accessed as it is done, e.g.,~in~\cite{CapelliI24,CarmeliTGKR23,BringmannCM22}. We then express this counting problem in terms of matrix multiplication where we have to compute the result of a product of a linear number of small matrices. To enable the binary search we then require that this product can be efficiently maintained under substitutions of some of the matrices. This type of problem is known in the literature as \emph{dynamic word problems} where the input is a word whose letters are elements from some semi-group whose product has to be maintained under element substitutions. Since we want to allow more powerful updates to the input later on, we cannot use the known data structures for dynamic word problems directly. We instead opt for a simpler approach that uses an extension of binary search trees to store our matrix product. The price of this is modest as it leads to a data structure that on only a doubly logarithmic factor slower than the provably optimal data structures for dynamic word problems. Plugging these ingredients together leads to a direct access algorithm with preprocessing time linear in the size of the input word and polylogarithmic access time. Moreover, the query to evaluate is given in the form of an unambiguous automaton---which is well known to be equivalent in expressivity to MSO-queries~\cite{buchi1960weak}---then all runtime dependencies on the query are polynomial.

One advantage of our search tree based data structure is that, by relying heavily on known tree balancing techniques, it allows updating the input string efficiently without running the preprocessing from scratch. Such update operations have been studied before in the enumeration literature for first-order logic~\cite{BerkholzKS18}, conjunctive queries~\cite{BerkholzKS17,BerkholzKS18} and, most intensively for MSO on words and trees~\cite{BalminPV04,LosemannM14,NiewerthS18,AmarilliBM18,Niewerth18,AmarilliBMN19b,KleestMeissnerMN22}. Compared to this latter line of work, we support more powerful updates: instead of single letter, respectively node, additions, and deletions as in most of these works, our data structure efficiently allows typical text operations like deletion of whole parts of the text, cut and paste, and concatenation of documents. The most similar update operations in the literature are those from~\cite{SchmidS22} which beside our operations allow persistence of the data structure and thus, in contrast to us, also copy and paste operations. Still, our updates are vastly more powerful than in all the rest of the literature. Moreover, we are the first to propose updates to direct access data structures; all works with updates mentioned above only deal with enumeration.

Another advantage of our algorithm is that it easily allows direct access to the query result for any lexicographic order. It has been observed that ranking the orders by a preference stated by the users is desirable. Thus, there are several works that tackle this question for direct access, enumeration, and related problems, see e.g.~\cite{BringmannCM22,CarmeliTGKR23,BakibayevKOZ13,DoleschalKMP22,DeepHK22,TziavelisGR20,TziavelisGR22}. 
While the lexicographic orders we consider are more restricted, %
we still consider them very natural. Also, it is interesting to see that, while there is an unavoidable runtime price to pay for ranked access with respect to lexicographic orders to conjunctive queries~\cite{CarmeliTGKR23,BringmannCM22}, we here get arbitrary lexicographic access for free by simply changing the order in which variables are treated during the binary search. As a consequence, we can even choose the order at access time as the preprocessing is completely agnostic to the order.

\paragraph{Outline of the paper} In Section~\ref{sec:preliminaries}, we start by introducing the formal notation and some basic results that will be used in the paper. In Section~\ref{sec:results}, we formalize the setting of direct access for MSO queries and state our main result. In Section~\ref{sec:matrices}, we show our matrix approach and algorithm for solving the direct access problem. In Section~\ref{sec:data_structure}, we show the data structure required for implementing the algorithm. In Section~\ref{sec:updates}, we study the direct access problem with edits and show how to extend our solution for this setting. Finally, in Section~\ref{sec:conclusions} we present some conclusions and future work.

This article is an extended version of~\cite{BourhisCMR25}, which includes all proofs omitted in the previous version.  
\cristian{I added the outline of the paper and a line stating what is new in this paper. If there is more new content, please, state it here.} 	
	\section{Preliminaries}\label{sec:preliminaries}

\paragraph{Sets and strings} For a set $A$, we denote by $2^A$ the power set of $A$. For $n \in \bbN$ with $n \geq 1$, we denote by $[n]$ the set $\{1, \ldots, n\}$. 
We use $\Sigma$ to denote a finite alphabet and $\Sigma^*$ all strings (also called words) with symbols in $\Sigma$. We will usually denote strings by $s, s', s_i$ and similar. For every $s_1, s_2\in \Sigma^*$, we write $s_1 \cdot s_2$ (or $s_1s_2$ for short) for the concatenation of $s_1$ and $s_2$. If $s = a_1 \ldots a_n \in \Sigma^*$, we write $|s| = n$, that is, the length of $s$. We denote by $\epsilon \in \Sigma^*$ the string of $0$ length (i.e., $|\epsilon| = 0$), also called the empty string. 
For $s= a_1 \ldots a_n\in \Sigma^*$, we denote by $s[i,j]$ the substring $a_i\ldots a_j$, by $s[..i]$ the prefix $a_1 \ldots a_i$ and by $s[i..]$ the suffix $a_i\ldots a_n$.

\paragraph{Mappings} Given a finite set $X$ of variables and $n \in \bbN$, in this work we will heavily work with \emph{mappings} of the form $\mu: X \rightarrow [n]$ as our outputs. We denote by $\dom{\mu} = X$ the \emph{domain} of~$\mu$ and by $\range{\mu} = \{i \in [n] \mid \exists x \in X\colon \, \mu(x) = i\}$ the \emph{range} of $\mu$.
Further, we denote by $\mu_\emptyset$ the \emph{empty mapping}, which is the unique mapping such that $\dom{\mu} = \emptyset$.
In our examples, we will usually write $(x_1 \mapsto i_1, \ldots, x_\ell \mapsto i_\ell)$ to denote the mapping $\mu: \{x_1, \ldots, x_\ell\} \rightarrow [n]$ such that $\mu(x_1) = i_1$, \ldots, $\mu(x_\ell) = i_\ell$. We also write $\mu \cup (x \mapsto i)$ with $x \notin \dom{\mu}$ for the mapping~$\mu'$ such that $\mu'(x) = i$ and $\mu'(y) = \mu(y)$ for every $y \in \dom{\mu}$. 

Given a mapping $\mu: X \rightarrow [n]$, we define $\mu^{-1}: [n] \rightarrow 2^X$ as the \emph{set-inverse mapping} of~$\mu$, defined by $\mu^{-1}(i) := \{x \in X \mid \mu(x) = i\}$.
Note that $i \in \range{\mu}$ if, and only if, $\mu^{-1}(i) \neq \emptyset$. 
Moreover, given a subset $Y \subseteq X$, we define the projection $\pi_{Y}(\mu)$ as the mapping $\mu': Y \rightarrow [n]$ such that $\mu'(y) = \mu(y)$ for every $y \in Y$. Remark that if $Y = \emptyset$, then $\pi_{Y}(\mu) = \mu_\emptyset$.

\paragraph{Variable-set automata} %
A \emph{variable-set automaton}~\cite{fagin2015document,florenzano2020efficient,AmarilliBMN19}
(or vset automaton for short) is a tuple
$
\cA = (Q, \Sigma, X, \Delta, q_0, F)
$
where $Q$ is a finite set of states, $X$ is a finite set of variables, $q_0 \in Q$ is an initial state, $F \subseteq Q$ is a finite set of states, and 
$
\Delta \ \subseteq \ Q \times \Sigma \times 2^X \times Q$ is the transition relation. 
Given a string $s = a_1 \ldots a_n$, a run $\rho$ of $\cA$ over $s$ is a sequence:
\begin{equation}
	\rho \ := \ q_0 \, \longtrans{a_1/X_1} \, q_1  \, \longtrans{a_2/X_2} \, \ldots \, \longtrans{a_n/X_n} \, q_n \label{eq:vset-run}
\end{equation}
such that $(q_{i-1}, a_i, X_i, q_{i+1}) \in \Delta$ for every $i \leq n$.
We say that a run $\rho$ like (\ref{eq:vset-run}) is \emph{valid} if, and only if, $\bigcup_{i=1}^{n} X_i = X$ and $X_i \cap X_j = \emptyset$ for every $i < j \leq n$; in other words, each variable in~$X$ appears exactly once in $\rho$. 
If $\rho$ is valid, one can define the mapping $\mu_\rho\colon X \rightarrow [n]$ such that, for every $x \in X$, $\mu_\rho(x) = i$ where $i$ is the unique position that satisfies $x \in X_i$. 
As usual, we also say that a run $\rho$ like (\ref{eq:vset-run}) is \emph{accepting} if, and only if, $q_n \in F$. 
Then we define the \emph{output} of $\cA$ over a string $s$ as the set of mappings:
\[
\sem{\cA}(s) \ = \ \{\mu_\rho \mid \text{$\rho$ is a valid and accepting run of $\cA$ over $s$}\}.
\]

We define a \emph{partial run} $\rho$ of $\cA$ as a sequence of transitions $\rho := p_0 \longtrans{b_1/Y_1} p_1 \longtrans{b_2/Y_2} \ldots \longtrans{b_n/Y_n} p_n$ such that $(p_{i-1}, b_i, Y_i, p_i) \in \Delta$. We say that $\rho$ is a partial run \emph{from $p_0$ to $p_n$}
\emph{over the string $b_1 \dots b_n$}. Note that a run is also a partial run where we additionally assumed that $p_0 = q_0$. We say that a partial run is valid if $Y_i\cap Y_j= \emptyset$ for all $i\le j\le n$.
We define the length of $\rho$ as $|\rho|=n$, and we make the convention that a single state $p_0$ is a partial run of length $0$. 
We define the set of variables $\vars(\rho)$ of a partial run $\rho$ as $\vars(\rho) = \bigcup_{i=1}^n Y_i$. 
Given two partial runs $\rho = p_0 \longtrans{b_1/Y_1}\ldots \longtrans{b_n/Y_n} p_n$ and $\sigma = r_0 \longtrans{c_1/Z_1} \ldots \longtrans{c_m/Z_m} r_m$ such that $p_n = r_0$ we define the run $\rho \cdot \sigma$ as the concatenation of $\rho$ and $\sigma$, i.e., the partial run: 
\[
\rho \cdot \sigma \ := \  p_0 \longtrans{b_1/Y_1} \ldots \longtrans{b_n/Y_n} p_n \longtrans{c_1/Z_1} \ldots \longtrans{c_m/Z_m} r_m.
\] 
Note that $|\rho \cdot \sigma| = |\rho| + |\sigma|$ and $\vars(\rho \cdot \sigma) = \vars(\rho) \cup \vars(\sigma)$. 

We define the size $|\cA$| of a vset automaton $\cA = (Q, \Sigma, X, \Delta, q_0, F)$ as the number of states and transitions, so $|\cA| := |Q| + |\Delta|$. 
In the following, we assume that all vset automata are \emph{trimmed}, i.e., for every $q \in Q$ there exists a run $\rho$ that reaches $q$, and there exists a partial run $\sigma$ from $q$ to some state in $F$. We can make this assumption without loss of generality, since removing unreachable states can be done in time $\cO(|\cA|)$ without modifying $\sem{\cA}$.  

\paragraph{MSO and vset automata} In this paper, we usually refer to \emph{Monadic Second Order Logic (MSO)} as our language for query evaluation; however, we will not define it formally here since we will use vset automata as an equivalent model for MSO. Precisely, when we refer to MSO, we mean MSO formulas of the form $\varphi(x_1, \ldots, x_n)$ over strings where $x_1, \ldots, x_n$ are first-order open variables (see, e.g.,~\cite{libkin2004elements} for a formal definition). Then, given a string~$s$ as a logical structure, the \emph{MSO query evaluation problem} refers to computing the set:
\[
\sem{\varphi(x_1, \ldots, x_n)}(s) \ := \  \{\mu: \{x_1, \ldots, x_n\} \rightarrow [|s|] \mid w, \mu \models \varphi(x_1, \ldots, x_n)\}
\] 
which is the set of all first-order assignments (i.e., mappings) that satisfy $\varphi$ over $s$. One can show that MSO over strings with first-order open variables is equally expressible to vset automata, basically, by following the same construction as for the Büchi-Elgot-Trakhtenbrot theorem~\cite{buchi1960weak} (see also~\cite{MunozR22}). Furthermore, vset automata (as defined here) are equally expressive to regular document spanners~\cite{fagin2015document} for information extraction (see, e.g., \cite{AmarilliJMR22,MunozR22}). 
For this reason, we can use vset automata to define MSO queries, which also applies to the setting of document spanners, or any other query language equally expressive to MSO. 

\paragraph{Functional and unambiguous vset automata} It is useful to consider vset automata that have no accepting runs that are not valid, so we make the following definition: we say that a vset automaton is \emph{functional} if, and only if, for every $s \in \Sigma^*$, every accepting run $\rho$ of $\cA$ over $s$ is also valid. In~\cite{fagin2015document}, it was shown that there is an algorithm that, given a vset automaton $\cA$, constructs in exponential time a functional vset automaton $\cA'$ of exponential size with respect to $\cA$ such that $\sem{\cA} = \sem{\cA'}$. We will thus restrict our analysis and algorithms to functional vset automata, since we can extend them to non-functional vset automata, incurring an exponential~blow-up.

One useful property of functional vset automata is that each state determines the variables that are assigned before reaching it in the following sense.
\begin{lem}\label{lemma:vars-q}
	Let $\cA = (Q, \Sigma, X, \Delta, q_0, F)$ be a functional vset automaton. For every $q \in Q$ there exists a set $X_q \subseteq X$ such that for every partial run $\rho$ from $q_0$ to $q$ it holds that $\vars(\rho) = X_q$. 
\end{lem}
\begin{proof}[Proof (sketch).]
	The lemma is standard in the literature for similar models, see e.g.~\cite{maturana2018document}. We provide a short proof. By way of contradiction, suppose that there exists a state $q \in Q$ and two runs $\rho$ and $\rho'$ that reach $q$ such that $\vars(\rho) \neq \vars(\rho')$. 
	Given that $\cA$ is trimmed, there exists a partial run $\sigma$ starting from $q$ that reaches some final state in $F$. 
	Then the runs $\rho \cdot \sigma$ and $\rho' \cdot \sigma$ are accepting. Since $\cA$ is functional, both runs are also valid and should satisfy $\vars(\rho \cdot \sigma) = X = \vars(\rho' \cdot \sigma')$. However, $\vars(\rho \cdot \sigma) \neq \vars(\rho' \cdot \sigma)$ which is a contradiction.  
\end{proof}

By \autoref{lemma:vars-q}, for every functional vset automaton $\cA = (Q, \Sigma, X, \Delta, q_0, F)$ and every $q \in Q$ we define its sets of variables as $X_q$ to be the set as in the lemma. One consequence of \autoref{lemma:vars-q} is that for every transition $(q, a, Y, q') \in \Delta$, we have $Y = X_{q'} \setminus X_q$. In particular, there can only be $|\Sigma|$ transitions for every pair $q, q'$ of states such that overall $|\Delta| = O(|Q|^2 \cdot \Sigma)$.%

We will also restrict to \emph{unambiguous} vset automata. We say that a vset automaton $\cA = (Q, \Sigma, X, \Delta, q_0, F)$ is \emph{unambiguous} if for every string $s \in \Sigma^*$ and every $\mu \in \sem{\cA}(s)$ there exists exactly one accepting run $\rho$ of $\cA$ over $s$ such that $\mu = \mu_\rho$. 
It is well-known that for every vset automaton $\cA$ one can construct an equivalent unambiguous functional vset automaton $\cA'$ of exponential size with respect to $\cA$.
Therefore, we can restrict our analysis to the class of unambiguous functional vset automaton without losing expressive power.
Both are standard assumptions in the literature for MSO query evaluation and have been used to find efficient enumeration algorithms for computing $\sem{\cA}(s)$, see e.g.~\cite{florenzano2020efficient,SchmidS21,MunozR22}.

\begin{exa}\label{ex:runs}

  \begin{figure} 
          
    \centering
    \begin{subfigure}[b]{0.3\linewidth}
      \includegraphics[width=.85\linewidth]{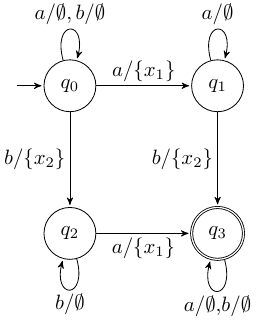}
      \caption{The vset automaton $\cA_0$. \vspace{3.5mm}}
      \label{fig:vsetautomata}
    \end{subfigure}
    \quad
    \begin{subfigure}[b]{0.33\linewidth}
      \centering
      \begin{tabular}{c c}
        $x_1$ & $x_2$ \\ \hline
        1 & 2 \\
        4 & 2 \\
        4 & 5 \\
        4 & 3 \\ 
      \end{tabular}
	\vspace{12mm}
      \caption{The mappings computed by $\cA_0$ on input string $s=abbab$.}
      \label{fig:relation}
    \end{subfigure}
\quad
    \begin{subfigure}[b]{0.3\linewidth}
      \centering
      \footnotesize
      
      $M_a = \begin{array}{c|cccc}
        & {q_0} & {q_1} & {q_2} & {q_3} \\ \hline
        {q_0} & 1 & 1 & 0 & 0 \\
        {q_1} & 0 & 1 & 0 & 0 \\
        {q_2} & 0 & 0 & 0 & 1 \\
        {q_3} & 0 & 0 & 0 & 1 
      \end{array}$
      \bigskip
      
      $M_b = \begin{array}{c|cccc}
            & {q_0} & {q_1} & {q_2} & {q_3} \\ \hline
        {q_0} & 1 & 0 & 1 & 0 \\
        {q_1} & 0 & 0 & 0 & 1 \\
        {q_2} & 0 & 0 & 1 & 0 \\
        {q_3} & 0 & 0 & 0 & 1 
      \end{array}$
      
      \vspace{5mm}

      \caption{Transition matrices of~$\cA_0$. \vspace{3.5mm}}
      
      \label{fig:matrices}
    \end{subfigure}
\caption{A running example of a vset automaton $\cA_0$ that will be used throughout this work.}%
    \label{fig:example}
  \end{figure}

  \autoref{fig:vsetautomata} depicts an unambiguous functional vset automaton $\cA_0$ on variables $\{x_1,x_2\}$ and alphabet $\{a,b\}$. It can be  verified that $X_{q_0} = \emptyset$, $X_{q_1} = \{x_1\}$, $X_{q_2} = \{x_2\}$ and $X_{q_3} = \{x_1,x_2\}$. Let $s = abbab$.  The following are all valid and accepting runs of $\cA_0$ over $s$:
\begin{itemize}
\item  $\rho_0 \ := \ q_0 \, \longtrans{a/\{x_1\}} \, q_1
                        \, \longtrans{b/\{x_2\}} \, q_3
                        \, \longtrans{b/\emptyset} \, q_3
                        \, \longtrans{a/\emptyset} \, q_3
                        \, \longtrans{b/\emptyset} \, q_3$ and $\mu_{\rho_0} = \{x_1 \mapsto 1,x_2 \mapsto 2\}$
\item  $\rho_1 \ := \ q_0 \, \longtrans{a/\emptyset} \, q_0
                        \, \longtrans{b/\{x_2\}} \, q_2
                        \, \longtrans{b/\emptyset} \, q_2
                        \, \longtrans{a/\{x_1\}} \, q_3
                        \, \longtrans{b/\emptyset} \, q_3$ and $\mu_{\rho_1} = \{x_1 \mapsto 4,x_2 \mapsto 2\}$
\item  $\rho_2 \ := \ q_0 \, \longtrans{a/\emptyset} \, q_0
                        \, \longtrans{b/\emptyset} \, q_0
                        \, \longtrans{b/\emptyset} \, q_0
                        \, \longtrans{a/\{x_1\}} \, q_1
                        \, \longtrans{b/\{x_2\}} \, q_3$ and $\mu_{\rho_2} = \{x_1 \mapsto 4,x_2 \mapsto 5\}$
\item  $\rho_3 \ := \ q_0 \, \longtrans{a/\emptyset} \, q_0
                        \, \longtrans{b/\emptyset} \, q_0
                        \, \longtrans{b/\{x_2\}} \, q_2
                        \, \longtrans{a/\{x_1\}} \, q_3
                        \, \longtrans{b/\emptyset} \, q_3$ and $\mu_{\rho_3} = \{x_1 \mapsto 4,x_2 \mapsto 3\}$
\end{itemize}

\noindent In \autoref{fig:relation}, we show a summary of $\sem{\cA_0}(s)$.
It can be verified that $\sem{\cA_0}(s)$ is the set of tuples $\mu$ such that $\mu(x_1)$ is a position where there is an $a$ in $s$, $\mu(x_2)$ is a position where there is a $b$ in $s$. Moreover, if $\mu(x_1) < \mu(x_2)$ then $\mu(x_2)$ is the first position after $\mu(x_1)$ containing a $b$. Otherwise, if $\mu(x_2) < \mu(x_1)$, then $\mu(x_1)$ is the first position after $\mu(x_2)$ containing an $a$.
\end{exa}

\paragraph{Computational model} We work with the usual RAM model with unit costs and logarithmic size registers, see e.g.~\cite{GrandjeanJ22}. %
All data structures that we will encounter in our algorithms will be of polynomial size, so all addresses and pointers will fit into a constant number of registers. Also, all atomic memory operations like random memory access and following pointers can be performed in constant time. The numbers we consider will be of value at most $|s|^{|X|}$ where $X$ is the variable set of the automaton we consider. Thus, all numbers can be stored in at most $|X|$ memory cells and arithmetic operations on them can be performed in time~$O(|X|^2)$ naively\footnote{We remark that there are better algorithms for arithmetic operations, but we will not follow this direction here. If the reader is willing to use faster algorithms for multiplication and addition, then in all runtime bounds below the factor $|X|^2$ can be be reduced accordingly.}. In the sequel, we use $\omega$ to denote the exponent for matrix multiplication.

	\section{Direct access for MSO queries}\label{sec:results}

Let $X$ be a set of variables. Given a total order $\prec$ on $X$, we extend $\prec$ to a \emph{lexicographic order} on mappings as usual: for mappings $\mu, \mu': X \rightarrow [n]$ we define $\mu \prec \mu'$ if, and only if, there exists $x \in X$ such that $\mu(x) < \mu'(x)$ and, for every $y \in X$, if $y \prec x$, then $\mu(y) = \mu'(y)$.  

Fix a total order $\prec$ on $X$. Given a vset automaton $\cA = (Q, \Sigma, X, \Delta, q_0, F)$ and an input string $s$, consider the set of outputs $\sem{\cA}(s) = \{\mu_1, \mu_2, \ldots\}$ such that $\mu_1 \prec \mu_2 \prec \ldots$.
We define the \emph{$i$-th output of $\cA$ over $s$}, denoted by $\sem{\cA}(s)[i]$, as the mapping~$\mu_i$. 
Intuitively, we see $\sem{\cA}(s)$ as an array where the outputs are ordered by $\prec$ and we retrieve the $i$-th element of this array. 
The \emph{direct access problem} for vset automata is the following: given a vset automaton $\cA$, a string $s$, and an index $i$, compute the $i$-th output $\sem{\cA}(s)[i]$; if the index~$i$ is larger than the number of solutions then return an out-of-bound error. 
Without loss of generality, in the following we always assume that $i \leq |\sem{\cA}(s)|$, since $|\sem{\cA}(s)|$ is easy to compute for unambiguous functional vset automata, see Section~\ref{sec:matrices}.
As usual, we split the computation into two phases, called the \emph{preprocessing} phase and the \emph{access} phase:
\bigskip
\begin{center}
	\framebox{
		\begin{tabular}{rl}
			\textbf{Problem:} \!\!\!\!\!\!& $\directaccessproblem$  \\ \hline \vspace{-3mm} \\
			\textbf{Preprocessing:}  \!\!\!\!\!\! & 
			$\left\{ \text{\begin{tabular}{rl}
					\textbf{input:} & \!\!\! a vset automaton $\cA$ and $s \in \Sigma^*$ \\
					\textbf{result:}
					& \!\!\! a data structure $D_{\cA,s}$
			\end{tabular}} \right.
			$            \\ \hline \vspace{-3mm} \\
			\textbf{Access:} \!\!\!\!\!\! &
			$\left\{
			\!\!\text{
				\begin{tabular}{rl}
					\textbf{input:} & \!\!\!
					\parbox[t]{4.2cm}{an index $i$ and $D_{\cA,s}$} \\
					\textbf{output:} & \!\!\! \parbox[t]{4.2cm}{the $i$-th
					 output $\sem{\cA}(s)[i]$}
				\end{tabular} 
			}\right.$\!\!\!\!
		\end{tabular}
	}
\end{center}
\bigskip
During the preprocessing phase, the algorithm receives as input a vset automaton $\cA$ and a string $s \in \Sigma^*$ and computes a data structure $D_{\cA, s}$. After preprocessing, there is the access phase where the algorithm receives any index $i$ as input and computes the $i$-th output $\sem{\cA}(s)[i]$ by using the precomputed data structure $D_{\cA, s}$. 

To measure the efficiency of a direct access algorithm, we say that an algorithm for the problem $\directaccessproblem$ has \emph{$f$-preprocessing} and \emph{$g$-access time} for some functions $f$ and $g$ if, and only if, the running time of the preprocessing phase and the access phase is in $O(f(\cA, s))$ and $O(g(\cA, s))$, respectively. 
Note that the running time of the access phase does not depend on $i$, given that its bitsize is bounded by $|X|\cdot \log(|s|)$ so that it can be stored in $|X|$ registers in the RAM model and arithmetic on it can be done in time $O(|X|^2)$~naively.

Given a class $\mathcal{C}$ of vset automata, %
we define the problem $\directaccessproblem$ for $\mathcal{C}$ as the problem %
above when we restrict $\cA$ to $\mathcal{C}$. The main result of this work is the~following.

\begin{thm}\label{thm:main}
 There is an algorithm that solves $\directaccessproblem$ for the class of unambiguous functional vset automata for any variable order $\prec$ with preprocessing $O(|Q|^\omega \cdot |X|^2 \cdot |s|)$ and access time $O(|Q|^\omega \cdot |X|^3  \cdot \log^2(|s|))$. These bounds remain even true when we assume that the order $\prec$ is only given as an additional input in the access phase.
\end{thm}

In \autoref{thm:main}, we restrict to the class of unambiguous vset automata. A natural question is what happens if we consider the class of all functional vset automata. We first remark that the \emph{data complexity} of the problem will not change as it is possible to translate any functional vset automaton to an unambiguous one with an exponential blow-up.
Unfortunately, in combined complexity, there is no polynomial time algorithm under standard assumptions.

\begin{prop} \label{theo:lowerbound}
 If $\directaccessproblem$ for the class of functional vset automata has an algorithm with polynomial time preprocessing and access, then {\sc P} $=$ {\sc NP}.
\end{prop}
  \begin{proof}
    The proof of this result exploits the fact that $\directaccessproblem$ allows answering the query $\sem{\cA}(s) \geq i$ for any index $i$. Indeed, when asking to access the $i$-th result, the algorithm will return an out-of-bounds error if $i$ is greater than $|\sem{\cA}(s)|$. By using the fact that $|\sem{\cA}(s)|$ is less than $|s|^{|X|}$, a binary search approach allows us to find  $|\sem{\cA}(s)|$ in $O(|\cA| \cdot \log |s|)$ steps by using $\directaccessproblem$ as an oracle.  By assuming that $\directaccessproblem$ has a polynomial time preprocessing and a polynomial time access time, we can conclude then that counting the number of solutions for a functional vset automaton is in {\sc FP}. Given that counting the number of solutions for functional vset automata is {\sc SpanL}-complete (Theorem 6.2 of \cite{florenzano2020efficient}), we can conclude that if $\directaccessproblem$ has a polynomial time preprocessing and a polynomial time access, then functions in {\sc SpanL} can be computed in {\sc FP}.

    {\sc SpanL} is the class of functions computable as $|R|$, where $R$ is the set of output values returned by the accepting paths of an {\sc NL} machine, see \cite{AlvarezJ93} for a formal definition. In \cite{AlvarezJ93}, it is shown that, if {\sc SpanL} $\subseteq$ {\sc FP}, i.e., every function in {\sc SpanL} is computable in polynomial time, then {\sc P} $=$ {\sc NP}. The statement of \autoref{theo:lowerbound} follows.
\end{proof}

\autoref{theo:lowerbound} conditionally shows that it is unlikely that there is an efficient algorithm for $\directaccessproblem$ for all functional vset automata in combined complexity. 

In the remainder of this paper, we will prove Theorem~\ref{thm:main}. To make the proof more digestible, in Section~\ref{sec:matrices} we show how to reduce the direct access problem to a counting problem and introducing a matrix approach for it. In Section~\ref{sec:data_structure}, we will present the data structure constructed during the preprocessing and used in the access phase, showing Theorem~\ref{thm:main}. In Section~\ref{sec:updates}, we then show how to integrate updates to the input string into our~approach.

	\section{From direct access to counting: a matrix approach}\label{sec:matrices}

In this section, we present the main algorithmic ideas for Theorem~\ref{thm:main}. In the next section, we use these ideas to develop a data structure for the preprocessing and access algorithm.

\paragraph{From direct access to counting} Let $\cA = (Q, \Sigma, X, \Delta, q_0, F)$ be a vset automaton and let $s = a_1 \ldots a_n$ be a string of length $n$. Assume that $X = \{x_1, \ldots, x_\ell\}$ has $\ell$ variables ordered by a total order $\prec$, w.l.o.g., $x_1 \prec x_2 \prec \ldots \prec x_{\ell}$. For an index $k \in [\ell]$ and a mapping $\tau: \{x_1, \ldots, x_k\} \rightarrow [n]$, we define the set:
\[
\sem{\cA}(s, \tau) = \{\mu \in \sem{\cA}(s) \mid \pi_{x_1, \ldots, x_{k-1}}(\mu) = \pi_{x_1, \ldots, x_{k-1}}(\tau) \ \wedge \ \mu(x_k) \leq \tau(x_k) \}.
\]
Intuitively, the set $\sem{\cA}(s, \tau)$ restricts the output set $\sem{\cA}(s)$ to all outputs in $\sem{\cA}(s)$ which coincide with $\tau$ for variables $x_i$ before the variable $x_k$, and that have a position before $\tau(x_k)$ for the variable~$x_k$. 
If $\tau = \mu_\emptyset$ is the empty mapping, we define $\sem{\cA}(s, \tau) = \sem{\cA}(s)$.

\begin{exa}
Going back to the example from \autoref{fig:vsetautomata} with $s = abbab$ and $\tau = (x_1 \mapsto 2)$ then $\sem{\cA_0}(s,\tau)$ contains all tuples from $\sem{\cA_0}(s)$ which map $x_1$ to a position before $2$. As seen  in \autoref{fig:relation} we have $\sem{\cA_0}(s, \tau) = \{(x_1 \mapsto 1, x_2 \mapsto 2)\} $. Now if $\tau = (x_1 \mapsto 2, x_2 \mapsto 3)$, we have $\sem{\cA_0}(s, \tau) = \emptyset$ since we only keep tuples where $x_1$ is set to $2$ and no such tuple can be found. If $\tau = (x_1 \mapsto 4, x_2 \mapsto 3)$, we have $\sem{\cA_0}(s, \tau) = \{(x_1 \mapsto 4, x_2 \mapsto 2), (x_1 \mapsto 4, x_2 \mapsto 3) \}$. 
\end{exa}

For the sake of presentation, we denote by $\cnt{\sem{\cA}(s, \tau)}$ the number $|\sem{\cA}(s, \tau)|$ of outputs in $\sem{\cA}(s, \tau)$. For direct access, we are interested in finding efficient algorithms for computing $\cnt{\sem{\cA}(s, \tau)}$ because of the following connection. 

\begin{lemC}[(Lemma 7 in \cite{CapelliI24})]
	\label{lemma:counting-to-access} 
	If there is an algorithm that computes $\cnt{\sem{\cA}(s, \tau)}$ in time~$T$ for every $k \in [\ell]$ and every $\tau:\{x_1, \ldots, x_k\} \rightarrow [n]$, then there is an algorithm that retrieves $\sem{\cA}(s)[i]$ in time $O(T \cdot \ell \cdot \log(n))$ for every index $i$.
\end{lemC}
\begin{proof}[Proof (sketch).]
	A similar proof can be found in \cite{CapelliI24}. For the convenience of the reader and because we will use a slightly more complicated variant later, we give a quick sketch here. Let $\mu_i = \sem{\cA}(s)[i]$. The idea is to first compute $\mu_i(x_1)$ by observing the following: $\mu_i(x_1)$ is the smallest value $j_1 \in [n]$ such that $\cnt{\sem{\cA}(s,(x_1 \mapsto j_1))} \geq i$. This value can be found in time $O(T \cdot \log n)$ by performing a binary search on $\{\cnt{\sem{\cA}(s,(x_1 \mapsto j_1))} \mid j_1 \leq n\}$. Once we have found $\mu_i(x_1)$, we compute $\mu_i(x_2)$ similarly by observing that it is the smallest value $j_2$ such that $\cnt{\sem{\cA}(s,(x_1 \mapsto j_1, x_2 \mapsto j_2))} \geq i-\cnt{\sem{\cA}(s,(x_1 \mapsto j_1-1))} $. The claim then follows by a simple induction.
\end{proof}
Given Lemma~\ref{lemma:counting-to-access}, in the following we concentrate our efforts on developing an index structure for efficiently computing $\cnt{\sem{\cA}(s, \tau)}$ for every $k \in [\ell]$ and every $\tau:\{x_1, \ldots, x_k\} \rightarrow [n]$.
We will achieve this goal by using a matrix representation of $\cA$ and reducing the problem to matrix multiplication. 

\paragraph{A matrix representation for the counting problem} Let $\cA = (Q,\Sigma, X, \Delta, q_0, F)$ be an unambiguous functional vset automaton. For any letter $a \in \Sigma$, we define the matrix $M_a \in \bbN^{Q \times Q}$ such that for every $p, q \in Q$:
\begin{align*}
M_a[p,q] \ := \ \left\{ 
\begin{array}{ll}
	1 & (p, a, S, q) \in \Delta \text{ for some $S \subseteq X$} \\ 
	0 & \text{otherwise.}
\end{array}
\right.
\end{align*}
Strictly speaking, the matrix $M_a$ depends on $\cA$ and we should write $M_a^\cA$; however, $\cA$ will always be clear from the context and, thus, we omit $\cA$ as a superscript from $M_a$. Since $\cA$ is functional, for every pair of states $p,q \in Q$ and $a \in \Sigma$ there exists at most one transition $(p, a, S, q) \in \Delta$. Then $M_a$ contains a $1$ for exactly the pairs of states that have a transition with the letter $a$. For this reason, one can construct $M_a$ in time $O(|Q|^2)$ for every $a \in \Sigma$.
\autoref{fig:matrices} gives an example of matrices $M_a$ and $M_b$ for the automaton $\cA_0$ from \autoref{ex:runs}. 

We can map strings to matrices in $\bbN^{Q \times Q}$ by homomorphically extending the mapping $a \mapsto M_a$ from letters to strings by mapping every string $s = a_1\ldots a_n$ to a matrix $M_s$ defined by the product $M_s := M_{a_1} \cdot M_{a_2} \cdot \ldots \cdot M_{a_n}$. For $\epsilon$, we define $M_\epsilon = I$ where $I$ is the identity matrix in $\bbN^{Q\times Q}$. Note that this forms an homomorphism from strings to matrices where $M_{s_1s_2} = M_{s_1} \cdot M_{s_2}$ for every pair of strings $s_1, s_2 \in \Sigma^*$. It is easy to verify that for all states $p,q\in Q$ we have that $M_s[p,q]$ is the number of partial runs from $p$ to $q$ of $\cA$ over $s$.
Furthermore, if we define $\vec{q}_0$ as the (row) vector such that $\vec{q}_0[p] = 1$ if $p = q_0$ and $0$, otherwise, and $\vec{F}$ as the (column) vector such that $\vec{F}[p] = 1$ if $p \in F$ and $0$ otherwise, then we have the following equality between number of outputs and matrix products:
\[
\cnt{\sem{\cA}(s)} = \vec{q}_0 \cdot M_s \cdot \vec{F}.
\]

Our goal is to have a similar result for calculating $\cnt{\sem{\cA}(s, \tau)}$ given a mapping $\tau: \{x_1, \ldots, x_k\} \rightarrow [n]$. For this purpose, for every string $s = a_1 \ldots a_n$, we define the matrix $M_s^\tau$ as follows. Recall that $\tau^{-1}$ is the set-inverse of $\tau$.
Then, for every $a \in \Sigma$ and $i \in [n]$, define the matrix $M_{a,i}^\tau \in \bbN^{Q \times Q}$ such that for every $p,q \in Q$:
\begin{align*}
M_{a,i}^\tau[p,q] \ = \ \left\{ 
\begin{array}{lll}
	1 & \tau^{-1}(i) \setminus \{x_k\} \subseteq S \text{ for some } (p, a, S, q) \in \Delta \text{ and } &   \ \ \ \hfill (1) \\
	& \text{if } \tau(x_k) = i \text{ then } x_k \in X_q; & \ \ \ \hfill (2) \\ 
	0 & \text{otherwise.}
\end{array}
\right.
\end{align*} 
Finally, we define $M^\tau_s = M_{a_1,1}^\tau \cdot M_{a_2,2}^\tau \cdot \ldots \cdot M_{a_n,n}^\tau$. 
Intuitively, Condition (1) for $M_{a,i}^\tau[p,q] = 1$ makes sure that all runs of $\cA$ over $s$ counted by $M^\tau_s$ must take transitions at position $i$ that contain all variables $x_{j}$ with $\tau(x_{j}) = i$ and $j < k$. Note that we remove $x_k$ from $\tau^{-1}(i)$ since $x_k$ has a special meaning in $\tau$. Indeed, Condition (2) for $M_{a,i}^\tau[p,q] = 1$ restricts $x_k$ such that its assignment must be before or equal to position $i$. For this second condition, we exploit the set $X_q$ of a functional vset automata, that gives us information of all variables that have been used until state $q$.

It is important to notice that if $i \notin \range{\tau}$ then $M^\tau_{a_i, i} = M_{a_i}$. Since $\range{\tau}$ has at most $k$ elements, the sequences $M_{a_1}, \ldots, M_{a_n}$ and $M_{a_1,1}^\tau, \ldots, M_{a_n,n}^\tau$ differ in at most $k$ matrices. In particular, if $\tau$ is the empty mapping, then $M^\tau_s = M_s$ as expected. 
Finally, similarly to $M_a$ one can compute $M_{a,i}^\tau$ in time $O(|Q|^2)$ for any $a \in \Sigma$ assuming that we have $\tau$ and~$\tau^{-1}$.

\begin{exa}
  Consider again the example from \autoref{fig:vsetautomata} with $s = abbab$. Observe that for any $\tau$, whenever $M_{a}[p,q]=0$ then necessarily $M_{a,i}^\tau[p,q]=0$. Hence in this example, we focus on example where the entry $M_{a}[p,q]$ goes from $1$ to $0$ after applying a partial mapping.  We let $\tau = (x_1 \mapsto 4, x_2 \mapsto 4)$. We have that $M^\tau_{a,4}[q_0,q_1] = 0$ because even if there is a transition $(q_0, a, S, q_1)$ with $x_1 \in S$, we do not have $x_2 \in X_{q_1}$.  In other words, a run compatible with $\tau$ cannot take transition $(q_0,a,\{x_1\},q_1)$ at position $4$ since it does not allow to map $x_2$ to a position before $4$. On the other hand, $M^\tau_{a,4}[q_2,q_3] = 1$  because it is possible to set $x_1$ to position $4$ by taking transition $(q_2, a, \{x_1\}, q_3)$. Moreover, since $x_2 \in X_{q_3}$, it means that $x_2$ has been necessarily set by an earlier transition, hence at a position preceding $4$. Finally, $M^\tau_{a,4}[q_3,q_3] = 0$ because there is no transition from $q_3$ to $q_3$ setting variable $x_1$.
\end{exa}

Similarly to the relation between $\cnt{\sem{\cA}(s)}$ and $M_s$, we can compute $\cnt{\sem{\cA}(s, \tau)}$ by using $M_s^\tau$ as the following result shows.

\begin{lem} \label{lemma:matrices-tau}
	For every unambiguous functional vset automaton $\cA = (Q,\Sigma, X, \Delta, q_0, F)$, every string $s \in \Sigma^*$, and every mapping $\tau: \{x_1, \ldots, x_k\} \rightarrow [n]$ it holds that:
	\[
	\cnt{\sem{\cA}(s, \tau)} = \vec{q}_0 \cdot M_s^\tau \cdot \vec{F}.
	\]
\end{lem}
\begin{proof}
	We fix a string $s$ of size $n$ and a mapping $\tau \colon \{x_1,\dots,x_k\} \rightarrow [n]$. We say that a run $\rho$ from $q_0$ to $q$ over a string $s'$ of length $n'$ is compatible with $\tau$ if for every $j \in [k-1]$ such that $\tau(x_j) \leq n'$, we have that $\mu_\rho(x_j)$ is defined and is such that $\mu_\rho(x_j) = \tau(x_j)$. Moreover, if $n' \geq \tau(x_k)$, we have that $\mu_\rho(x_k)$ is defined and $\mu_\rho(x_k) \leq \tau(x_k)$.
	
	Define $P_i := \Pi_{j=1}^i M_{a_i,i}^\tau$. We now prove by induction on $i$ that for any state $q$, the value $P_i[q_0,q]$ is the number of runs from $q_0$ to $q$ over the string $a_1 \dots a_i$ compatible with~$\tau$.
	
	The base case is for $i=1$. First assume that $\tau(x_k) \neq 1$. Then there is a run from $q_0$ to~$q$ over the string $a_1$ compatible with $\tau$ if, and only if, there is a transition $(q_0,a_1,S,q)$ such that $\tau^{-1}(1) \setminus \{x_k\} \subseteq S$. Moreover, observe that if such a transition exists, it has to be unique since otherwise the automaton would not be functional. In this case, $M_{1,a_1}^\tau[q_0,q]=1$ by definition, which is indeed the number of runs from $q_0$ to $q$ compatible with $\tau$. Now if $\tau(x_k) = 1$, then there is a run from $q_0$ to $q$ over the string $a_1$ compatible with $\tau$ if, and only if, there is a transition $(q_0,a_1,S,q)$ such that $\tau^{-1}(1) \setminus \subseteq S$. Now, if such a transition exists, then $x_{k} \in X_q$ and hence, $M_{1,a_1}^\tau[q_0,q] = 1$, which again, is the number of runs from $q_0$ to $q$ compatible with $\tau$.
	
	Now, assume that the claim holds for $i < n$.  By definition $P_{i+1} = P_i \cdot M_{a_{i+1},i+1}^\tau$, so:
	\begin{equation}
		P_{i+1}[q_0,q] = \sum_{q' \in Q} P_i[q_0,q'] M_{a_{i+1},i+1}^\tau[q',q].  \label{eq:recmat}
	\end{equation}
	for every $q$.
	By induction, for every $q'$, $P_i[q_0,q']$ is the number of run from $q_0$ to $q'$ compatible with $\tau$ over the string $a_1\dots a_i$. Observe that a run from $q_0$ to $q$ compatible with $\tau$ over string $a_1 \dots a_{i+1}$ is a run from $q_0$ to some state $q'$ compatible with $\tau$ followed by a transition $(q', a_{i+1}, S, q)$. To be compatible with $\tau$, the last transition should verify that $\tau^{-1}(i+1) \subseteq S$. Observe that by functionality, given $q'$, at most one such transition exists. Moreover, if $\tau(x_k) \geq i+1$, then the run has to map $x_k$ to a position before $i+1$. But this happens if, and only if, $x_k \in X_q$.
	
	We hence have two cases. If $\tau(x_k) \geq i+1$, then either $x_k \notin X_q$ and there are no run from $q_0$ to $q$ over $a_1\dots a_{i+1}$ compatible with $\tau$. But then $M_{a_{i+1}, i+1}^\tau[q',q] = 0$ for every $q'$ by definition and hence, $P_{i+1}[q_0,q] = 0$ by \autoref{eq:recmat} and the hypothesis holds. If $x_k \in X_q$, then the set of runs from $q_0$ to $q$ is the disjoint union over all $q'$, of the set of runs from $q_0$ to $q'$, followed by the transition $(q',a_{i+1},S,q)$ with $\tau^{-1}(i+1) \subseteq S$ when it exists. This transition exists if, and only if, $M_{a_{i+1},i+1}^\tau[q',q] = 1$ (recall we assumed $x_k \in X_q$). Hence, following \autoref{eq:recmat}, the number of runs from $q_0$ to $q'$ compatible with $\tau$ is indeed $P_{i+1}[q_0,q]$.
	
	Now if $\tau(x_k) \neq i+1$, then as before, the set of runs from $q_0$ to $q$ is the disjoint union, for every $q'$, of the set of runs from $q_0$ to $q'$, followed by the transition $(q',a_{i+1},S,q)$ with $\tau^{-1}(i+1) \subseteq S$ when it exists. This transition exists if, and only if, $M_{a_{i+1},i+1}^\tau[q',q] = 1$ (since $\tau(x_k) \neq i+1$). Hence, following \autoref{eq:recmat}, the number of runs from $q_0$ to $q'$ over $a_1 \dots a_{i+1}$ compatible with $\tau$ is indeed $P_{i+1}[q_0,q]$, which concludes the induction.
	
	We can now conclude the proof by observing that $\cnt{\sem{\cA}(s)}$ is the number of runs from $q_0$ to some final state $q \in F$ over $s$ compatible with $\tau$. Hence, it is $\sum_{q \in F} P_n[q_0,q]$ by what precedes, which is $\vec{q}_0 \cdot M_s^\tau \cdot \vec{F}$.
\end{proof}

\paragraph{The algorithm} Now that we have defined the main technical tools, we can present the algorithm for the preprocessing and direct access of an unambiguous functional vset automaton~$\cA$ over a string $s = a_1 \ldots a_n$. For this purpose, we will assume here the existence of a data structure for maintaining the product of matrices $\{M_{a_i,i}^\tau\}_{i \in [n]}$ to then present and analyze the algorithms. 
Intuitively, the data structure will maintain the product of $\{M_{a_i,i}^\tau\}_{i \in [n]}$ in a well-balanced binary tree and, for each update of a matrix $M_{a_i,i}^\tau$, we will update the product by traversing a branch of such tree. 
In Section~\ref{sec:data_structure}, we will explain this data structure in full detail.
Next, we define the interface of such data structure to then explain and analyze the main algorithm.

Suppose the existence of a data structure $D$ that given values $n, m \in \bbN$ maintains $n$ square matrices $M_1, \ldots, M_n$ of size $m\times m$. This data structure supports three methods called $\dinit$, $\dset$, and $\dout$ which are defined as follows:
\begin{itemize}
	\item $D \gets \dinit(M_1, \ldots, M_n)$: receive as input the initial $n$ matrices $M_1, \ldots, M_n$ of size $m \times m$ and initialize a data structure $D$ storing the matrices $M_1, \ldots, M_n$;
	\item $D' \gets \dset(D, j, M)$: receive as input a data structure $D$, a position $j \leq n$, an $m \times m$-matrix $M$, and output a data structure $D'$ that is equivalent to $D$ but its $j$-th matrix $M_j$ is replaced by $M$; and
	\item $M \gets \dout(D)$: receive as input a data structure $D$ storing matrices $M_1,  \ldots, M_n$ and outputs a pointer to the matrix $M := M_1 \cdot \ldots \cdot M_n$, i.e., the product of the $n$ matrices. 
\end{itemize} 
Further, we assume that method $\dset$ is \emph{persistent}~\cite{driscoll1986making}, meaning that each call to $\dset$ produces a new data structure $D'$ without modifying the previous data structure $D$. %

\begin{algorithm}[t]
	\caption{The preprocessing and direct access algorithms of an unambiguous functional vset automaton $\cA = (Q, \Sigma, X, \Delta, q_0, F)$ with variables $x_1, \ldots, x_\ell$ and string $s = a_1 \ldots a_n$.}\label{alg:final}
	\smallskip
	\begin{varwidth}[t]{0.5\textwidth}
		\begin{algorithmic}[1]
			\Procedure{{Preprocessing}}{$\cA$, $s$}
			\State $D \gets \dinit(M_{a_1}, \ldots, M_{a_n})$
			\EndProcedure
			\smallskip
			
			\Procedure{{BinarySearch}}{$x, i, \tau$}
			\State $L \gets 0$
			\State $R \gets n$
			\While{$L \neq R$}
			\State $j \gets \left\lceil(L+R)/2\right\rceil$
			\State $\tau' \gets \tau \cup (x \mapsto j)$
			\State $D' \gets \dset(D, j, M^{\tau'}_{a_j,j})$ 
			\If{$\vec{q}_0 \cdot \dout(D') \cdot \vec{F} \geq i$}
			\State $R \gets j$
			\Else
			\State $L \gets j$
			\EndIf
			\EndWhile
			\State \Return $R$
			\EndProcedure
			\algstore{myalg}
		\end{algorithmic}
	\end{varwidth} \hfill
	\begin{varwidth}[t]{0.5\textwidth}
		\begin{algorithmic}[1]
			\algrestore{myalg}
			\Procedure{{DirectAccess}}{$i$}
			\State $\tau \gets \mu_\emptyset$
			\For{$k = 1, \ldots, \ell$}
			\State $j \gets \textsc{BinarySearch}(x_k, i, \tau)$
			\State $i \gets i-\textsc{CalculateDiff}(x_k, j, \tau)$
			\State $\tau \gets \tau \cup (x_k \mapsto j)$
			\State $\textsc{UpdateStruct}(x_{k+1}, j, \tau)$
			\EndFor
			\State \Return $\tau$
			\EndProcedure
			\smallskip
			
			\Procedure{{CalculateDiff}}{$x$, $j$, $\tau$}
			\State $\tau_{\text{prev}} \gets \tau \cup (x \mapsto j-1)$
			\State $D_{\text{prev}} \gets \dset(D, j, M^{\tau_{\text{prev}}}_{a_{j-1}, j-1})$
			\State \Return $\vec{q}_0 \cdot \dout(D_{\text{prev}}) \cdot \vec{F}$
			\EndProcedure
			\smallskip
			
			\Procedure{{UpdateStruct}}{$x$, $j$, $\tau$}
			\State $\tau_{\text{next}} \gets \tau \cup (x \mapsto n+1)$
			\State $D \gets \dset(D, j, M^{\tau_{\text{next}}}_{a_j, j})$ 
			\EndProcedure
		\end{algorithmic}
	\end{varwidth}  \hfill
\end{algorithm}

Assuming the existence of the data structure $D$, in Algorithm~\ref{alg:final} we present all steps for the preprocessing and direct access of an unambiguous functional vset automaton $\cA$ over a~string $s = a_1 \ldots a_n$. For the sake of presentation, we assume that $\cA$, $s$, and the data structure $D$ are available globally by all procedures.

Algorithm~\ref{alg:final} uses all the tools developed in this section for solving the MSO direct access problem. The preprocessing (Algorithm~\ref{alg:final}, left) receives as input the vset automaton and the string and constructs the data structure $D$ by calling the method $\dinit$ with matrices $M_{a_1}, \ldots, M_{a_n}$. The direct access (Algorithm~\ref{alg:final}, right) receives any index $i$ and outputs the $i$-th mapping $\tau$ in $\sem{\cA}(s)$ by following the strategy of Lemma~\ref{lemma:counting-to-access}. Specifically, starting from an empty mapping $\tau$ (line 16), it finds the positions for variables $x_1, \ldots, x_\ell$ by binary search for each variable $x_k$ (line 16-17). After finding the value $j$ for $x_k$, it decreases the index $i$ by calculating the difference of outputs just before $j$ (line 19), updates $\tau$ with $(x_k\mapsto j)$ (line~20), and updates $D$ with the new value of $x_k$ (line 21). We use the auxiliary procedures $\textsc{CalculateDiff}$ and $\textsc{UpdateStruct}$ to simplify the presentation of these steps. The workhorse of the direct access is the procedure $\textsc{BinarySearch}$. It performs a standard binary search strategy for finding the value for variable $x$, by using the reduction of Lemma~\ref{lemma:counting-to-access} to the counting problem and the matrix characterization $\cnt{\sem{\cA}(s, \tau)}$ of Lemma~\ref{lemma:matrices-tau}.

The correctness of Algorithm~\ref{alg:final} follows from Lemma~\ref{lemma:counting-to-access}~and~\ref{lemma:matrices-tau}.
Regarding the running time, suppose that methods $\dinit$, $\dset$, and $\dout$ of the data structure $D$ take time $t_\dinit$, $t_\dset$, and $t_\dout$, respectively. For the preprocessing, one can check that the running time is $O(|Q|^2 \cdot |s| + t_\dinit)$ where it takes $O(|Q|^2 \cdot |s|)$ for creating the matrices $M_{a_1}, \ldots, M_{a_n}$. For the direct access, one can check that it takes time $O(|X| \cdot \log (|s|) \cdot (t_{\dset} + t_{\dout}))$ for each index $i$.  

The next section shows how to implement the data structure $D$ and its methods $\dinit$, $\dset$, and $\dout$. In particular, we show that the running time of these methods will be $t_{\dinit} = O(|Q|^\omega\cdot |X|^2 \cdot |s|)$, $t_{\dset} = O(|Q|^\omega \cdot |X|^2 \cdot \log(|s|))$, and $t_{\dout} = O(1)$. %
Overall, the total running time of Algorithm~\ref{alg:final} will be $O(|Q|^\omega\cdot |X|^2 \cdot |s|)$ for the preprocessing and $O(|Q|^\omega \cdot |X|^3 \cdot \log(|s|)^2)$ for each direct access as stated in Theorem~\ref{thm:main}.
 	
	\section{Maintaining semi-groups products}\label{sec:data_structure}

In this section, we will show how to implement the data structure required in Section~\ref{sec:matrices}, which in particular has a $\dset$ operation that allows to change some of the elements in a sequence of matrices while maintaining their product efficiently. Similar problems have been studied under the name of \emph{dynamic word problems} in the literature, see e.g.~\cite{FrandsenMS97,AmarilliJP21}. In that setting, one is given a sequence of elements from a semi-group and wants to maintain the product of this sequence under substitution of the elements. Depending on the algebraic structure of the semi-group, there are algorithms of different efficiencies, and there is by now quite a good understanding, see again~\cite{FrandsenMS97,AmarilliJP21}.

We could likely use the results of~\cite{FrandsenMS97} directly to define our data structure (though~\cite{FrandsenMS97} assumes a finite semiring which does not correspond to our setting). %
However, later we will want to support more powerful changes to the strings than just substitutions, and it is not clear how the approach from~\cite{FrandsenMS97} could be adapted for this. So we choose to use a less technically involved approach that allows for more powerful update operations while only losing a factor of $\log\log(n)$ compared to~\cite{FrandsenMS97}. So fix a semi-group $\SG$, i.e.,~a set $\SG$ with an associative operation $\circ$. The semi-group that will be of most interest to us are the $(m\times m)$-matrices with elements in $\mathbb{N}$ with the usual matrix multiplication, but our approach here works for any semi-group. As it is common, we often leave out the multiplication symbol $\circ$ and write $g_1g_2$ for $g_1\circ g_2$. We next introduce our data~structure.

We will store sequences $\vec{g}=g_1,\ldots, g_n$ over $\SG$ in binary trees. To this end, let $T=(V,E)$ be a rooted binary tree in which the vertices are labeled with elements from $\SG$. We assume that the children of each internal node of $T$ are either a left or a right child. The label of a vertex $v$ is denoted by $\lab(v)$. Remember that the in-order traversal of a rooted binary tree, when encountering a vertex $v$ with left child $v_1$ and right child $v_2$, first recursively visits the subtree of $v_1$, then $v$, and finally recursively the subtree $v_2$. For every vertex $v$ of $T$, we define the sequence of $v$, in symbols $\word(v)$, to be the sequence of elements in the subtree rooted in $v$ read in in-order traversal. We also define the \emph{product} of $v$, in symbols $\prod(v)$ as the product of the elements in $\word(v)$ in the order of the sequence. Note that $\prod(v)$ is a single element from $\SG$ while $\word(v)$ is a sequence of elements from $\SG$; we never store $\word(v)$ explicitly in the vertex $v$ but store $\prod(v)$. We also write $\word(T)$ for the sequence of the root of $T$ and analogously define $\prod(T)$. %

In the remainder, we want to maintain $\prod(T)$ under changes to $\word(T)$. To this end, it will be useful to access positions in $\word(T)$ efficiently. If the only changes were substitutions of elements as needed for Algorithm~\ref{alg:final}, we could easily do this by letting $T$ be a binary search tree in which the keys are the position in~$\word(T)$. However, we later also want to be able to delete, add, and move strings, and for these operations we would have to change a linear number of keys which is too expensive for our setting. Instead, we use the linear list representation with search-trees, see e.g.~\cite[Chapter 6.2.3]{Knuth98a}: we only store the \emph{size} of the subtree of $T$ rooted in $v$, which we denote by $\size(v)$. Remark that $\size(v)$ is also the length of $\word(v)$.
The $\size$ information lets us locate the element at desired positions in~$\word(v)$.

\begin{obs}\label{obs:descent}
    For every vertex $v$ in $T$ we can, given an index $i\in [|\word(v)|]$ and correct $\size$-labels, decide in constant time if the entry of position $i$ in $\word(v)$ is in the left subtree, the right subtree or in $v$ itself. In the two former cases, we can also compute at which position it is in the contiguous subsequence of that subtree.
\end{obs}

\autoref{obs:descent} directly gives an easy algorithm that descends from the root of $T$ to any desired position in $\word(T)$ in time linear in the height of $T$.

As the final information, we also add, for each vertex $v$ of $T$, the height of $v$, denoted by $\height(v)$. This information is not needed in this section, but it will be useful in~\autoref{sec:updates} for maintaining balanced trees, so we use it already here.
Since the height of $T$ will determine the runtime of our algorithm, we aim to bound it as for binary search trees.
We say that a tree $T$ satisfies the \emph{AVL condition} if for every vertex $v$ with children $v_1, v_2$ we have that $|\height(v_1)- \height(v_2)| \le 1$ and if $v$ has only one child $v_1$, then $v_1$ is a leaf. The AVL condition allows us to bound the height of trees by the following classical result.

\begin{thmC}[\cite{AdelsonVL62}]
    Every tree of size $n$ that satisfies the AVL condition has height $O(\log(n))$.
\end{thmC}

We now have everything to define our data structure: we call a tree $T$ in which all vertices are labeled with elements from $\SG$, which have correct attributes $\size$, $\height$, and $\prod$ as above and which satisfies the AVL-condition an \emph{AVL-product representation of $\word(T)$}.

\begin{figure}
	\centering
	\includegraphics[width=12cm]{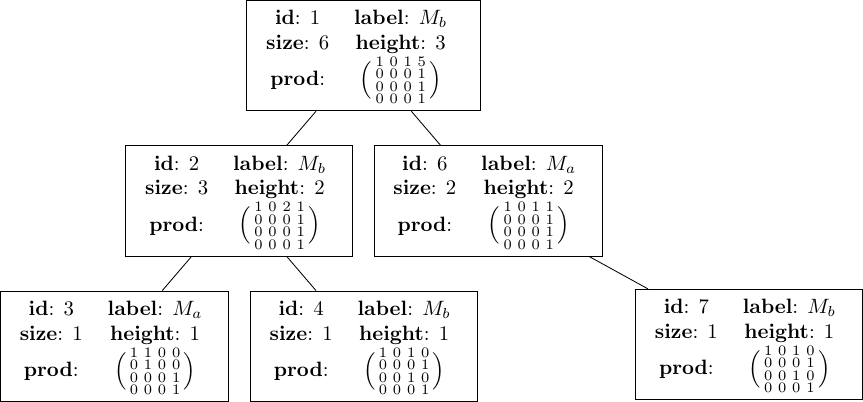}
	\caption{AVL-product representation for the string $abbbab$ and automaton $\cA_0$ from \autoref{fig:vsetautomata}.}
	\label{fig:avltree}
\end{figure}

\begin{exa}
  We consider again the automaton $\cA_0$ from \autoref{fig:vsetautomata} and string $abbbab$ (this is a different string than the one considered in previous examples as a longer string makes illustration more relevant for the data structure). In this case, we show an AVL-product representation of $M_aM_bM_bM_bM_aM_b = M^{\mu_\emptyset}_{a,1}M^{\mu_\emptyset}_{b,2}M^{\mu_\emptyset}_{b,3}M^{\mu_\emptyset}_{b,4}M^{\mu_\emptyset}_{a,5}M^{\mu_\emptyset}_{b,6}$ in \autoref{fig:avltree}. 
  For convenience, despite this not being mandatory, we assume each vertex $v$ of the tree also is identified by a unique number $\id(v)$. In the example, we can verify that $\prod$ values are computed from the labels. For example, if we denote by $v_i$ the vertex having $\id(v_i) = i$, we have $\prod(v_2) = \prod(v_3) \cdot M_b \cdot \prod(v_4)$ which recursively is equal to $M_aM_bM_b$. Similarly, $\prod(v_1) = \prod(v_2) \cdot \prod(v_6)$, that is, $M_aM_bM_bM_bM_aM_b$, which is the product we want to maintain.
\end{exa}

We now show that AVL-product representations are useful in our setting. We start by observing that $\dout$-queries are trivial since we store the necessary information $\prod(T)$ explicitly in the root.

\begin{obs}
    Given an AVL-product representation $T$, one can return $\prod(T)$ in constant time. So in particular, if $\word(T)$ is a sequence of matrices, AVL-product representations implement the method $\dout$ in constant time.
\end{obs}

We next show that there is an efficient initialization algorithm.

\begin{lem}\label{lemma:init}
    There is an algorithm that, given a sequence $\vec{g}$ over $\SG$ of length $n$, computes in time $O(n)$ and with $O(n)$ semi-group products an AVL-product representation of $\vec{g}$.
  \end{lem}
  \begin{proof}
	We first compute the maximal integer $d$ such that $2^{d+1}-1\leq n$. Then we construct a full, balanced binary tree $T'$ of height $d$ and thus with $2^{d+1}-1$ vertices in time $O(n)$. We then traverse $T'$ and add $n-(2^{d+1}-1)$ leaves to end up with a tree $T$ of height $d+1$ with $n$ vertices. Note that we can add enough leaves in the process since $T'$ has $2^d$ leaves, so we can add up to $2^{d+1}$ new leaves, ending up with $2^{d+1}-1+2^{d+1} = 2^{d+2}-1 > n$ vertices overall. By construction, $T$ satisfies the AVL-condition.
	
	We then add the labels $\lab(v)$ along an in-order traversal. Finally, we compute $\size(v)$, $\height(v)$ and $\prod(v)$ along a post-order traversal, so a traversal that first visits the children of any vertex $v$ before $v$ itself. Each traversal takes time $O(n)$, plus $O(n)$ operations for the labels and $O(n)$ semi-group operations for the $\prod$-labels.
\end{proof}

Specializing to matrices, we get the following for the $\dinit$-operation of Algorithm~\ref{alg:final}.

\begin{cor}
    We can perform $\dinit$ for AVL-product representations in time~$O(m^{\omega}n)$.
\end{cor}

Finally, we show that AVL-product representations can be used for the substitution updates that we need for direct access when using the $\dset$ operation in a persistent manner.

\begin{lem}\label{lemma:maintainProduct}
    There is an algorithm that, given an AVL-product representation $T$, a position $i$ and a semi-group element $g$, outputs a new AVL-product representation of the sequence we get from $\word(T)$ by substitution the element in position $i$ by $g$. The runtime is $O(\log(|\word(T)|))$ plus the time to compute $O(\log(|\word(T)|))$ products in the semi-group.
\end{lem}
\begin{proof}
  We use \autoref{obs:descent} to descend from the root of $T$ to the vertex $v$ containing in $\lab(v)$ the element we want to substitute. To make this operation persistent, during the descent, we construct a copy of the path (called as \emph{path-copying}~\cite{driscoll1986making}) from the root to $v$ where the copy $u'$ of every vertex $u$ has all the same labels the original $u$ and points to the same child vertices not on the subtree. We then change $\lab(v')$ to the desired value and propagate the right $\prod$-values along bottom up along the path from $v'$ to the root $r'$. The tree rooted in $r'$ is the desired AVL-product representation and can be computed in time $O(\height(T)) = O(\log(n))$ and the same number of semi-group operations.
\end{proof}
      
\begin{cor}
    The $\dset$-method from Algorithm~\ref{alg:final} can be implemented in time $O(m^\omega \log(n))$ with AVL-product representations.
\end{cor}
\begin{exa}
  Going back to the example from \autoref{fig:avltree}. Assume now that one wants to update the data structure to only keep runs where $x_1$ is set to a position lower or equal than~$2$. To this end, one needs to compute $M^{\tau}_{a,1}M^{\tau}_{b,2}M^{\tau}_{b,3}M^{\tau}_{b,4}M^{\tau}_{a,5}M^{\tau}_{b,6} = M^{\mu_\emptyset}_{a,1}M^{\tau}_{b,2}M^{\mu_\emptyset}_{b,3}M^{\mu_\emptyset}_{b,4}M^{\mu_\emptyset}_{a,5}M^{\mu_\emptyset}_{b,6}$ where $\tau = (x_1 \mapsto 2)$. Hence, we only have to change one matrix in the original product.

  To do so, going down in the tree using $\size(\cdot)$, we can identify that the vertex where the second matrix is introduced in the product is vertex $v_2$. Hence, we only have to update $\prod(v_2)$ and $\prod(v_1)$. To do so, observe that the new value $M_2'$ of $\prod(v_2)$ is $\prod(v_3) \cdot M_{b,2}^\tau \cdot \prod(v_4)$, which can be obtained with two matrix products, $\prod(v_3)$ and $\prod(v_4)$ being already computed. Similarly, the new value of $\prod(v_1)$ is $M_2'\cdot  M_b \cdot  \prod(v_6)$ which can also be computed with two extra matrix products, provided that we compute $M_2'$ first. Hence, in general, this update scheme can be performed with at most $2 \cdot  \height(T)$ matrix products.
\end{exa}

    \section{Dynamic direct access under complex editing operations}\label{sec:updates}

In the previous sections, we have established that, given a string $s \in \Sigma^*$ of length $n$ and an unambiguous functional automaton $\cA$, we can answer direct access queries on the relation $\sem{\cA}(s)$ in time $O(\log(n)^2)$ after $O(n)$ preprocessing. Now, assume that the string $s$ is modified. 
If one wants to perform direct access on the relation induced by $\cA$ on the updated string
so far one has to perform the initialization of \autoref{lemma:init} again which will have $O(n)$ complexity for each editing step. In this section, we will show that the AVL-product representations from \autoref{sec:data_structure} can be updated efficiently so that we can perform direct access 
on the edited string
without redoing the preprocessing from scratch. 
We show that we can even support more complex edits than just appending letters. In a more likely scenario, the user can simultaneously work with two (or more) strings, performing complex operations over them. 

In the following, we formalize the scenario of complex edits over strings. Then we define the problem of dynamic MSO direct access. We finish by showing how to easily extend our data structure and approach to give a solution for this scenario.

\paragraph{Editing programs over strings} Fix an alphabet $\Sigma$. Let $\sdb = \{s_1, \ldots, s_N\}$ be a set of strings over $\Sigma$ called a \emph{strings database}. For the sake of simplicity, we use $s_i \in \sdb$ both as a string and as a \emph{string name} (i.e., as a label referring to $s_i$ that is considered of constant size). Let $\sset$ be a set of \emph{string variables} $S_1, S_2, \ldots$ disjoint from $\sdb$ and $\Sigma$. 
A \emph{literal} $l$ is either a string name $s_i \in \sdb$, a string variable $S \in \sset$, or a symbol $a \in \Sigma$.
An \emph{editing rule} is a command of the following four types:
\[
S := \econcat(l, l') \ \ \ \ \ (S,S') := \esplit(l, i) \ \ \ \ \  (S,S') := \ecut(l, i, j)  \ \ \ \ \  S := \epaste(l, l', i) 
\]
where $S$ and $S'$ are string variables, $l$ and $l'$ are literals, and $i, j \in \bbN$. Intuitively, string variables will be assigned to strings and then $\econcat$ will allow to concatenate two strings, $\esplit$ to split a string at position $i$, $\ecut$ to extract the substring between positions $i$ and $j$, and $\epaste$ to insert a string inside another at position $i$. An \emph{editing program} $\Pi$ is a pair $(S, P)$ where $S \in \sset$ is the output string variable and $P$ is a sequence of editing rules $R_1; R_2; \ldots; R_n$ such that each string name appears at most once in the right-hand side of some rule, and each string variable appears at most once in the right-hand side of a rule $R_i$ after it appears in the left-hand side of a rule $R_j$ with $j < i$.
In other words, each string in the database can be used only once and each variable can be used at most once after it is defined. We define the size of an editing program $\Pi$ as the number of rules $|\Pi| = n$.

\begin{exa}\label{ex:edit-program}
	Suppose that we have a strings database $\sdb_0 = \{s_1, s_2\}$. We define the editing program $\Pi_0 = (S_6, P_0)$ with $P_0$ as the sequence:
	\[
	\begin{array}{rrcl}
		P_0:\ \ \ \  & (S_1, S_2) & := & \esplit(s_1, 4); \\
		& (S_3, S_4) & := & \esplit(S_2, 1); \\
		& S_5 & := & \econcat(S_1, a); \\ 
		& S_6 & := & \econcat(S_5, S_4).
	\end{array}
	\] 
\end{exa} 

Next, we define the semantics of editing programs. %
A \emph{string assignment} is a partial function $\sigma\colon (\sdb \cup \Sigma \cup \sset) \rightarrow  \Sigma^*$ such that $\sigma(a) = a$ for every $a \in \Sigma$ and  $\sigma(s) = s$ for every $s \in \sdb$. %
We define the trivial assignment $\sigma_{\sdb}$  such that $\dom{\sigma_{\sdb}} = \sdb \cup \Sigma$ (i.e., no variable is mapped). 
Given an assignment $\sigma$, a string variable $S$, and a string $s \in \Sigma^*$, we write $\sigma[S \mapsto s]$ to denote the assigment $\sigma'$ that replaces $S$ with $s$ in $\sigma$ (i.e., $\sigma'(S) = s$ and $\sigma'(S') = \sigma(S')$ for every $S' \in \dom{\sigma} \setminus \{S\}$). Then, we define the semantics of rules as a function $\sem{\cdot}$ that maps assignments to assignments such that for every assignment $\sigma$:
\[
\renewcommand{\arraystretch}{1.4}
\begin{array}{rcl}
	\sem{S := \econcat(l, l')}(\sigma) & = & \sigma[S \mapsto \sigma(l) \cdot \sigma(l')] \\
	\sem{(S,S') := \esplit(l, i)}(\sigma) & = & \sigma[S \mapsto \sigma(l)[..i],S' \mapsto \sigma(l)[i+1..]] \\
	\sem{(S,S') := \ecut(l, i, j)}(\sigma) & = & \sigma[S \mapsto \sigma(l)[i,j], S' \mapsto \sigma(l)[..i-1] \cdot \sigma(l)[j+1..]] \\
	\sem{S := \epaste(l, l', i)}(\sigma) & = & \sigma[S \mapsto \sigma(l)[..i] \cdot \sigma(l') \cdot \sigma(l)[i+1..]]
\end{array}
\]
where we use the syntax $\sigma[S \mapsto s, S' \mapsto s']$ as a shorthand for $(\sigma[S \mapsto s])[S' \mapsto s']$.
We extend the semantics to sequences of editing rules $R_1; \ldots; R_n$ recursively as follows:
\[
\sem{R_1; \ldots; R_{n-1}; R_n}(\sigma) \ = \ \sem{R_n}(\sem{R_1; \ldots; R_{n-1}}(\sigma)).
\]
Finally, we define the \emph{semantics of an editing program} $\Pi = (S, P)$ as the string:
\[
\sem{\Pi}(\sdb) \ = \  [\sem{P}(\sigma_{\sdb})](S)
\] namely, the string at $S$ after evaluating $P$ starting from the trivial assignment. We assume that the above semantics is undefined if any of the previous definitions are undefined like, for example, indices $i, j$ are out of range, the variable $S$ is not defined, etc. Note that one can easily check in time $O(|\Pi|)$ whether the semantics will be undefined or not. For this reason, we assume that the semantics of an editing program is always defined. 

\begin{exa}[continue]
	If we take $s_1 = bbbbcb$, we get $\sem{\Pi_0}(\sdb) = bbbbab$. In particular, one can easily check that the program $\Pi_0$ is updating the fifth letter of $s_1$ with a letter $a$.  
\end{exa} 

Our editing programs can easily simulate the insertion, deletion, or update of a symbol in a string. Further, they can simulate the complex document editing operations presented in~\cite{SchmidS22} like concat, extract, delete, and insert (see~\cite{SchmidS22} for a definition of these operations). So, our set of operations can be seen as nearly equivalent to the one presented in~\cite{SchmidS22}; the only operation that we disallow is copying. Indeed, we do not allow editing programs to make copies of the input strings or the string variables by definition. An editing program that makes copies could produce strings of exponential size, and the input index $i$ for direct access will be of polynomial size with respect to the input documents, breaking our assumptions regarding the RAM model. For this reason, we disallow copying in our programs and leave this extension for future work.

\paragraph{Dynamic direct access of MSO} Now that the notion of editing programs is clear, we can define the problem of dynamic direct access as an extension of the $\directaccessproblem$ introduced in Section~\ref{sec:results}, adding a phase of edits over the strings database. 
\bigskip
\begin{center}
	\framebox{
		\begin{tabular}{rl}
			\textbf{Problem:} \!\!\!\!\!\!& $\dyndirectaccessproblem$  \\ \hline \vspace{-3mm} \\
			\textbf{Preprocessing:}  \!\!\!\!\!\! & 
			$\left\{ \text{\begin{tabular}{rl}
					\textbf{input:} & \!\!\! a vset automaton $\cA$ and \\ 
					& a strings database $\sdb = \{s_1, \ldots, s_N\}$ \\
					\textbf{result:}
					& \!\!\! a data structure $D_{\cA,\sdb}$
			\end{tabular}} \right.
			$            \\ \hline \vspace{-3mm} \\
			\textbf{Editing:}  \!\!\!\!\!\! & 
			$\left\{ \text{\begin{tabular}{rl}
					\textbf{input:} & \!\!\! an editing program $\Pi$ and $D_{\cA,\sdb}$ \\
					\textbf{result:}
					& \!\!\! a data structure $D_{\Pi,\cA,\sdb}$
			\end{tabular}} \right.
			$            \\ \hline \vspace{-3mm} \\
			\textbf{Access:} \!\!\!\!\!\! &
			$\left\{
			\!\!\text{
				\begin{tabular}{rl}
					\textbf{input:} & \!\!\!
					\parbox[t]{4.2cm}{an index $i$ and $D_{\Pi,\cA,\sdb}$} \\
					\textbf{output:} & \!\!\! \parbox[t]{7cm}{the $i$-th output $\sem{\cA}(\sem{\Pi}(\sdb))[i]$}
				\end{tabular} 
			}\right.$\!\!\!\!
		\end{tabular}
	}
\end{center}
\bigskip
Contrary to the standard direct access, we add an intermediate phase, called editing, where one can receive an editing program $\Pi$ over $\sdb$ as input. Thus, the direct access is now performed over the resulting string $\sem{\Pi}(\sdb)$. 
As before, we measure each phase separately and say that an algorithm for $\dyndirectaccessproblem$ has $f$-preprocessing, $h$-editing, and $g$-access time for some functions $f$, $h$, and $g$ if, and only if, the running time of the preprocessing, editing, and access phases is in $O(f(\cA, \sdb))$, $O(h(\cA, \sdb, \Pi))$, and $O(g(\cA, \sdb, \Pi))$, respectively.

By taking advantage of our techniques for MSO direct access, we can show that $\dyndirectaccessproblem$ can be solved with logarithmic running time during the editing phase and without increasing the running time of the preprocessing and access~phases.

\begin{thm}\label{thm:main-dyn}
	There is an algorithm that solves $\dyndirectaccessproblem$, for the class of unambiguous functional vset automata for any variable order $\prec$ with preprocessing time $O(|Q|^\omega \cdot |X|^2 \cdot \sum_{i=1}^N|s_i|)$, editing time $O(|Q|^\omega \cdot |X|^2 \cdot |\Pi| \cdot \max_{i=1}^N \log(|s_i|))$, and access time $O(|Q|^\omega \cdot |X|^3 \cdot \log^2(|\Pi| \cdot \max_{i=1}^N|s_i|))$. These bounds remain even true when we assume that the order $\prec$ is only given as an additional input in the access phase.
\end{thm}

The proof goes by showing how to implement each editing rule receiving as inputs AVL-products representations. Precisely, the preprocessing constructs an AVL-product representation for each string $s_i$ in $\sdb$ (e.g., like in Section~\ref{sec:data_structure}). Then, during the editing phase we produce new AVL-product representations from the initial ones for each editing operation. Finally, the resulting representation for the string $\sem{\Pi}(\sdb)$ is used for the direct access phase. These give the desired running time for the preprocessing and the direct access phases. Therefore, it is only left to show how to implement each editing rule in time $O(|Q|^\omega \cdot |X|^2 \cdot  \max_{i=1}^N \log(|s_i|))$ in order to prove the running time for the editing phase.  

\paragraph{Implementing the editing rules}
It is easy to see that $(S,S') := \ecut(l, i, j)$ can be implemented by splitting twice (at $i$ and $j$) and one concatenation. More formally, as the sequence of editing rules $(S_1, S_2) := \esplit(l,j); \ (S_3,S) := \esplit(S_1, i); \ S' \gets \econcat(S_3, S_2)$. Similarly, $S:=\epaste(l,l',i)$ can be obtained by one split and two concatenations. More formally, $(S_1,S_2):= \esplit(l,i);\ S_3 := \econcat(S_1, l');\ S := \econcat(S_3, S_2)$. Finally, we can easily replace a single symbol by a newly initialized data structure. 
For this reason, we dedicate this subsection to show how to implement $\econcat$ and $\esplit$ efficiently.

Assume that the inputs of $\econcat$ and $\esplit$ are given by AVL-product representations. We will heavily rely on the corresponding operations on AVL trees as presented in~\cite{BlellochFS16}: there is it shown that many operations on ordered sets can be implemented with AVL trees by using only an operation called $\join$ that does the following: given two AVL trees $T_1, T_2$ and a key $k$, such that the maximal key in $T_1$ is smaller than $k$ and the minimal key in $T_2$ is bigger than $k$, $\join$ returns an AVL tree $T$ containing all keys in $T_1$, $T_2$ and $k$. Besides the usual tree navigation, the only basic operation that $\join$ uses is $\node$, which takes a key $k$ and two trees $T_1, T_2$ and returns a tree that has $k$ in its root and $T_1$ and $T_2$ as left and right subtrees, respectively\footnote{We remark that the usual rotations in search trees can be simulated by a constant number of $\node$-operations.}. We directly get the following result.

\begin{lem}\label{lemma:join}
	There is an algorithm $\join$ that, given two AVL-product representations $T_1, T_2$ and a semi-group element $g$, computes in time $O(\log(\size(T_1) +\size(T_2)))$ and the same number of semi-group operations an AVL-product representation $T$ of the sequence we get by concatenating $\word(T_1)$, $g$, and $\word(T_2)$.
\end{lem}
\begin{proof}
	We first use a variant of the $\join$ operation for AVL trees from~\cite{BlellochFS16}, see below for a description. The only difference to \cite{BlellochFS16} is that we never search for and compare keys of a search tree but instead positions in $\word(T_1)$ and $\word(T_2)$, respectively. This can easily be done with \autoref{obs:descent}, as before. To get a valid AVL-product representation, we only have to show how the labels for $\prod$ and $\size$ can be computed correctly. However, since the only way we ever change the structure of any tree in the $\join$-algorithm is by the $\node$-operator, we can simply compute the correct values from those of the child nodes whenever we apply that operator. The runtime bound then comes directly from that of the join operation for AVL trees.

\begin{algorithm}[t]
    \caption{The algorithm for the $\join$ operation of AVL-trees}
    \label{alg:join}
    \begin{algorithmic}[1] %
        \Procedure{joinRight}{$T_L,k,T_R$}
        \State $(\ell,k', c) \gets \Call{expose}{T_L}$
        \If{$\height(c) \le \height(T_R)+1$} \!\Comment{We are at the position where $T_R$ can be~inserted.}
            \State $T' \gets \Call{Node}{c,k,T_r}$
            \If{$\height(T')\le \height(\ell)$} 
            \State \Return $\Call{Node}{\ell, k', T'}$
            \Else
                \State \Return $\Call{rotateLeft}{}(\Call{Node}{}(\ell, k', \Call{rotateRight}{T'}))$
            \EndIf
        \Else \Comment{Have to descend recursively into the tree to find height to insert $T_R$.}
            \State $T' \gets \Call{joinRight}{c, k, T_R}$
            \State $T''\gets \Call{Node}{\ell, k', T'}$
            \If{$\height(T') \le \height(\ell)+1$ }
            \State \Return $T''$
            \Else
            \State \Return $\Call{rotateLeft}{T''}$
            \EndIf
        \EndIf
        \EndProcedure
        \Procedure{join}{$T_L,k,T_R$} %
        \If{$\height(T_L)>\height(T_R)+1$} 
        \State \Return \Call{joinRight}{}$(T_L, k, T_R)$
        \EndIf
        \If{$\height(T_R)>\height(T_L)+1$} 
        \State \Return \Call{joinLeft}{}$(T_L, k, T_R)$
        \EndIf
        \State \Return $\Call{Node}{T_L, k, T_R}$ \Comment{Input trees roughly same height, so simply combine} 
\EndProcedure
    \end{algorithmic}
\end{algorithm}

In Algorithm~\ref{alg:join}, we give the implementation of the join operation for AVL trees as in~\cite{BlellochFS16}. The idea is as follows: if the input trees are of the same height (up to a difference of $1$), we can simply combine them and are done. Otherwise, we have to join the less high tree into the higher tree for which we have two functions $\textsc{joinRight}$ and $\textsc{joinLeft}$. We here only give the code for $\textsc{joinRight}$ since $\textsc{joinLeft}$ is totally symmetric. The idea is to recursively descend into the higher tree from the root to a level in which the less high tree can be inserted. We do so, and then have to go up and might apply some rotations to fix the AVL condition that might be broken by the insertion of the new tree. In Algorithm~\ref{alg:join}, we use several building blocks for which we do not give the code: $\Call{Node}{T_1, k, T_2}$ returns a tree that consists of $k$ in the root, left subtree $T_1$ and right subtree $T_2$. Inverse to this is $\Call{expose}{T}$ which, given a tree with root $r$, left subtree $T_1$ and right subtree $T_2$, returns the triple $(T_1, r, T_2)$. Finally, $\textsc{rotateLeft}$ and $\textsc{rotateRight}$ are the usual rotation operations in binary tree that can be implemented by a combination of a constant number of calls to $\textsc{expose}$ and $\textsc{Node}$. Note that $\textsc{Node}$ implicitly also computes the correct value for $\height$, so we can assume that $\height$ is always correct in our algorithms.

It can be shown that $\textsc{join}$ takes time proportional to the length of the descent into the higher tree, so $O(|\height(T_L)- \height(T_R)|)$, see~\cite{BlellochFS16}. 
\end{proof}

\begin{algorithm}[t]
    \caption{The algorithm for the $\join$ operation of AVL-trees}
    \label{alg:split}
    \begin{algorithmic}[1] %
        \Procedure{split}{$T,k$}
        \State $(L,m,R) \gets \Call{expose}{T}$
        \If{$k=m$} \Comment{Found target key, can return its subtrees.}
        \State \Return $(L,R)$
        \ElsIf{$k < m$}\Comment{Have to descent into left subtree.}
        \State $(L_L, L_R) \gets \Call{split}{L,k}$
        \State \Return $(L_L, \Call{join}{L_R, R})$
        \Else \Comment{Have to descent into right subtree.}
        \State $(R_L, R_R) \gets \Call{split}{R,k}$
        \State \Return $(\Call{join}{L, R_L}, R_R)$
        \EndIf
        \EndProcedure
    \end{algorithmic}
\end{algorithm}

Using $\join$, one can implement an operation called $\splita$ that takes as input an AVL tree and a key $k$ and returns two AVL trees $T_1$ and $T_2$ such that all keys in $T$ smaller than~$k$ are in $T_1$ while all keys bigger than $k$ are in $T_2$. Besides $\join$, the only operation that $\splita$ performs is a descent in the tree to find the node with key $k$. 
This directly gives the following algorithms for operations on $\word(T)$.

\begin{lem}\label{lemma:split}
	There is an algorithm $\splita$ that, given an AVL-product representation $T$ and a position $i$, computes in time $O(\log(\size(T)))$ and the same number of semi-group operations two trees AVL-product representations $T_1, T_2$ such that $\word(T_1)$ is the prefix of $\word(T)$ up to but excluding the $i$-th entry and $\word(T_2)$ is the suffix of $\word(i)$ that starts after the $i$-th~position.
\end{lem}
\begin{proof}
	We use the $\splita$ operation for AVL-trees, but again instead of explicit keys using positions with \autoref{obs:descent}. By definition, this computes the trees with the right words. Since the only operations that change any trees in $\splita$ are calls of $\join$, we automatically get that the result are AVL-product representations in which all labels are correct. The runtime bound comes directly from that for $\splita$ on AVL trees and the fact that we only changed the implementation of $\node$ in the proof of \autoref{lemma:join} which only leads to a constant overhead and one semi-ring operation in every call.

For $\textsc{split}$, we give the algorithm in Algorithm~\ref{alg:split}.\footnote{Compared to~\cite{BlellochFS16}, we give a simplified version in which we assume that the key that we are looking for is always in the tree. This is safe because in our application we will consider as keys that positions in the~word.} The idea is that we descend from the root into the tree until we find the key that we are looking for. During the descent, we remember which subtrees contain elements bigger and smaller than the key $k$, respectively, and use the join operation to collect them in two corresponding trees.

The runtime of $\textsc{split}$ is proportional to the number of steps the algorithm has to descend until it find the key, so in the worst case $O(\log(n))$, see~\cite{BlellochFS16}.
A key observation that we use in our AVL-product representations is that all steps that change the input tree are only $\textsc{Node}$ operations which is why we only have to consider them in our proofs.
\end{proof}

Applying both lemmas for matrices, we directly get bounds for $\econcat$ and $\esplit$.
\begin{cor} \label{cor:impl-edits}
	There are algorithms for both $S:= \econcat(l, l')$ and $(S,S'):=\esplit(l,i)$ on AVL-product representations that, given inputs $T_1, T_2$, resp., $T$ and $i$, run in time $O(m^\omega\cdot \log(|T_1|+|T_2|))$, resp., $O(m^\omega \cdot \log(|T|))$.
\end{cor}
\begin{proof}
	For $S:= \econcat(l, l')$, we first use $\splita$ to take out the last element $s$ of the product and call the resulting AVL-product representation $T_1'$. Then we use $\join(T_1', s, T_2)$ to compute the desired result.
	
	For $(S,S'):=\esplit(l,i)$, we first use $\splita(T,i)$ to get AVL-product representations, remembering the element $s$ at position $i$. Call $T_1, T_2$ the outputs of this operation. Then we split of the element $s'$ at position $i-1$ from $T_1$ and compute an AVL-product representation~$T_s$ containing only $s$. Finally, compute $T_1'$ by the call $\join(T_1', s', T_s)$. The desired output is then $(T_1', T_2)$. 
	
	All runtime bounds follow directly from \autoref{lemma:join}, \autoref{lemma:split}, and \autoref{lemma:init}.
\end{proof}

	\section{Conclusions and future work}\label{sec:conclusions}

\cristian{I added the previous long future work section that we have before. Please check.}

We have given a direct-access algorithm for MSO queries on strings that supports powerful updates and all lexicographic orders without requiring them during preprocessing.  Arguably,  the algorithm is simple and uses, in particular, no complicated combinatorial constructions or deep automata-theoretic results. Instead, a combination of matrix multiplication, binary search, and a slight extension of binary search trees give a flexible and efficient algorithm. 

Our result is certainly only a first step toward understanding direct access for MSO queries, which motivates new research questions. First, it would be interesting to see to what extent our results can be adapted to MSO on trees. It is plausible to design an efficient direct access algorithm in that setting by balancing the tree, say via a tree decomposition as in~\cite{AmarilliBM18} or the forest algebra approach of~\cite{KleestMeissnerMN22}, and then using our matrix multiplication technique on this balanced representation. While this would provide efficient direct access, the update operations in that case would likely be quite limited, allowing only lead addition and deletion, as in~\cite{KleestMeissnerMN22}, and node relabeling. For more powerful updates, one would need to determine whether there is an operation similar to \textsc{join} for AVL trees. Another issue is that balancing distorts the tree's structure, so it is not clear whether any natural tree-structure-based orders (such as the lexicographic orders here) could be supported by this approach.

Another interesting problem is to study if our approach can be adapted to support copy-and-paste, arguably one of the most common text operations. To do so, as in~\cite{SchmidS22}, one would likely have to make our data structure persistent, i.e., the update operations would have to be non-destructive and leave the old version intact. Such algorithms are known for AVL trees~\cite{driscoll1986making}; however, we have not studied whether this is directly useful in our setting.

Another question is whether the access time of our algorithm can be reduced. Most direct access algorithms have logarithmic delay (see, e.g.,~\cite{CarmeliTGKR23}), and the delay of most enumeration algorithms is also at most logarithmic. So, it would be relevant to improve the access time by a logarithmic factor here.

One could also try to determine whether direct access to words (and trees) is possible for more expressive queries, e.g., the grammar-based extraction formalisms of~\cite{Peterfreund21,AmarilliJMR22}. For these, there are known enumeration algorithms, but we do not know if those extend to direct access.

Our last two questions concern the link to dynamic word problems, so the maintenance of the product over semi-groups is only for letter substitutions. It is known that the optimal update time is in general~$\Theta(\log(n)/\log\log(n))$~\cite{FrandsenMS97}, so a doubly logarithmic factor faster than our search-tree based implementation. Can this factor also be saved for our more general update operations? Also, there are even faster algorithms for dynamic word problems over some restricted types of  semi-groups~\cite{FrandsenMS97,AmarilliJP21}. Can we have these same runtimes also for more powerful updates than single-letter substitutions? We leave these questions for future work.

	\bibliographystyle{alphaurl}
	\bibliography{./extras/biblio2}	

\newcommand{\etalchar}[1]{$^{#1}$}
\begin{thebibliography}{ABMN19b}

\bibitem[ABM18]{AmarilliBM18}
Antoine Amarilli, Pierre Bourhis, and Stefan Mengel.
\newblock Enumeration on trees under relabelings.
\newblock In Benny Kimelfeld and Yael Amsterdamer, editors, {\em 21st
  International Conference on Database Theory, {ICDT} 2018, Vienna, Austria,
  March 26-29, 2018}, volume~98 of {\em LIPIcs}, pages 5:1--5:18. Schloss
  Dagstuhl - Leibniz-Zentrum f{\"{u}}r Informatik, 2018.
\newblock URL: \url{https://doi.org/10.4230/LIPIcs.ICDT.2018.5}, \href
  {https://doi.org/10.4230/LIPICS.ICDT.2018.5}
  {\path{doi:10.4230/LIPICS.ICDT.2018.5}}.

\bibitem[ABMN19a]{AmarilliBMN19}
Antoine Amarilli, Pierre Bourhis, Stefan Mengel, and Matthias Niewerth.
\newblock Constant-delay enumeration for nondeterministic document spanners.
\newblock In Pablo Barcel{\'{o}} and Marco Calautti, editors, {\em 22nd
  International Conference on Database Theory, {ICDT} 2019, Lisbon, Portugal,
  March 26-28, 2019}, volume 127 of {\em LIPIcs}, pages 22:1--22:19. Schloss
  Dagstuhl - Leibniz-Zentrum f{\"{u}}r Informatik, 2019.
\newblock URL: \url{https://doi.org/10.4230/LIPIcs.ICDT.2019.22}, \href
  {https://doi.org/10.4230/LIPICS.ICDT.2019.22}
  {\path{doi:10.4230/LIPICS.ICDT.2019.22}}.

\bibitem[ABMN19b]{AmarilliBMN19b}
Antoine Amarilli, Pierre Bourhis, Stefan Mengel, and Matthias Niewerth.
\newblock Enumeration on trees with tractable combined complexity and efficient
  updates.
\newblock In Dan Suciu, Sebastian Skritek, and Christoph Koch, editors, {\em
  Proceedings of the 38th {ACM} {SIGMOD-SIGACT-SIGAI} Symposium on Principles
  of Database Systems, {PODS} 2019, Amsterdam, The Netherlands, June 30 - July
  5, 2019}, pages 89--103. {ACM}, 2019.
\newblock \href {https://doi.org/10.1145/3294052.3319702}
  {\path{doi:10.1145/3294052.3319702}}.

\bibitem[{\`{A}}J93]{AlvarezJ93}
Carme {\`{A}}lvarez and Birgit Jenner.
\newblock A very hard log-space counting class.
\newblock {\em Theor. Comput. Sci.}, 107(1):3--30, 1993.
\newblock \href {https://doi.org/10.1016/0304-3975(93)90252-O}
  {\path{doi:10.1016/0304-3975(93)90252-O}}.

\bibitem[AJMR22]{AmarilliJMR22}
Antoine Amarilli, Louis Jachiet, Martin Mu{\~{n}}oz, and Cristian Riveros.
\newblock Efficient enumeration for annotated grammars.
\newblock In Leonid Libkin and Pablo Barcel{\'{o}}, editors, {\em {PODS} '22:
  International Conference on Management of Data, Philadelphia, PA, USA, June
  12 - 17, 2022}, pages 291--300. {ACM}, 2022.
\newblock \href {https://doi.org/10.1145/3517804.3526232}
  {\path{doi:10.1145/3517804.3526232}}.

\bibitem[AJP21]{AmarilliJP21}
Antoine Amarilli, Louis Jachiet, and Charles Paperman.
\newblock Dynamic membership for regular languages.
\newblock In Nikhil Bansal, Emanuela Merelli, and James Worrell, editors, {\em
  48th International Colloquium on Automata, Languages, and Programming,
  {ICALP} 2021, Glasgow, Scotland (Virtual Conference), July 12-16, 2021},
  volume 198 of {\em LIPIcs}, pages 116:1--116:17. Schloss Dagstuhl -
  Leibniz-Zentrum f{\"{u}}r Informatik, 2021.
\newblock URL: \url{https://doi.org/10.4230/LIPIcs.ICALP.2021.116}, \href
  {https://doi.org/10.4230/LIPICS.ICALP.2021.116}
  {\path{doi:10.4230/LIPICS.ICALP.2021.116}}.

\bibitem[AVL62]{AdelsonVL62}
Georgii~M Adel'son-Vel'skii and Evgenii Landis.
\newblock An algorithm for the organization of information.
\newblock {\em Soviet Math.}, 3:1259--1263, 1962.

\bibitem[BCM22]{BringmannCM22}
Karl Bringmann, Nofar Carmeli, and Stefan Mengel.
\newblock Tight fine-grained bounds for direct access on join queries.
\newblock In Leonid Libkin and Pablo Barcel{\'{o}}, editors, {\em {PODS} '22:
  International Conference on Management of Data, Philadelphia, PA, USA, June
  12 - 17, 2022}, pages 427--436. {ACM}, 2022.
\newblock \href {https://doi.org/10.1145/3517804.3526234}
  {\path{doi:10.1145/3517804.3526234}}.

\bibitem[BCMR25]{BourhisCMR25}
Pierre Bourhis, Florent Capelli, Stefan Mengel, and Cristian Riveros.
\newblock Dynamic direct access of {MSO} query evaluation over strings.
\newblock In Sudeepa Roy and Ahmet Kara, editors, {\em 28th International
  Conference on Database Theory, {ICDT} 2025, Barcelona, Spain, March 25-28,
  2025}, volume 328 of {\em LIPIcs}, pages 26:1--26:18. Schloss Dagstuhl -
  Leibniz-Zentrum f{\"{u}}r Informatik, 2025.
\newblock URL: \url{https://doi.org/10.4230/LIPIcs.ICDT.2025.26}, \href
  {https://doi.org/10.4230/LIPICS.ICDT.2025.26}
  {\path{doi:10.4230/LIPICS.ICDT.2025.26}}.

\bibitem[BDGO08]{BaganDGO08}
Guillaume Bagan, Arnaud Durand, Etienne Grandjean, and Fr{\'{e}}d{\'{e}}ric
  Olive.
\newblock Computing the jth solution of a first-order query.
\newblock {\em {RAIRO} Theor. Informatics Appl.}, 42(1):147--164, 2008.
\newblock URL: \url{https://doi.org/10.1051/ita:2007046}, \href
  {https://doi.org/10.1051/ITA:2007046} {\path{doi:10.1051/ITA:2007046}}.

\bibitem[BFS16]{BlellochFS16}
Guy~E. Blelloch, Daniel Ferizovic, and Yihan Sun.
\newblock Just join for parallel ordered sets.
\newblock In Christian Scheideler and Seth Gilbert, editors, {\em Proceedings
  of the 28th {ACM} Symposium on Parallelism in Algorithms and Architectures,
  {SPAA} 2016, Asilomar State Beach/Pacific Grove, CA, USA, July 11-13, 2016},
  pages 253--264. {ACM}, 2016.
\newblock \href {https://doi.org/10.1145/2935764.2935768}
  {\path{doi:10.1145/2935764.2935768}}.

\bibitem[BKOZ13]{BakibayevKOZ13}
Nurzhan Bakibayev, Tom{\'{a}}s Kocisk{\'{y}}, Dan Olteanu, and Jakub Zavodny.
\newblock Aggregation and ordering in factorised databases.
\newblock {\em Proc. {VLDB} Endow.}, 6(14):1990--2001, 2013.
\newblock URL: \url{http://www.vldb.org/pvldb/vol6/p1990-zavodny.pdf}, \href
  {https://doi.org/10.14778/2556549.2556579}
  {\path{doi:10.14778/2556549.2556579}}.

\bibitem[BKS17]{BerkholzKS17}
Christoph Berkholz, Jens Keppeler, and Nicole Schweikardt.
\newblock Answering conjunctive queries under updates.
\newblock In Emanuel Sallinger, Jan~Van den Bussche, and Floris Geerts,
  editors, {\em Proceedings of the 36th {ACM} {SIGMOD-SIGACT-SIGAI} Symposium
  on Principles of Database Systems, {PODS} 2017, Chicago, IL, USA, May 14-19,
  2017}, pages 303--318. {ACM}, 2017.
\newblock \href {https://doi.org/10.1145/3034786.3034789}
  {\path{doi:10.1145/3034786.3034789}}.

\bibitem[BKS18]{BerkholzKS18}
Christoph Berkholz, Jens Keppeler, and Nicole Schweikardt.
\newblock Answering {FO+MOD} queries under updates on bounded degree databases.
\newblock {\em {ACM} Trans. Database Syst.}, 43(2):7:1--7:32, 2018.
\newblock \href {https://doi.org/10.1145/3232056} {\path{doi:10.1145/3232056}}.

\bibitem[BPV04]{BalminPV04}
Andrey Balmin, Yannis Papakonstantinou, and Victor Vianu.
\newblock Incremental validation of {XML} documents.
\newblock {\em {ACM} Trans. Database Syst.}, 29(4):710--751, 2004.
\newblock \href {https://doi.org/10.1145/1042046.1042050}
  {\path{doi:10.1145/1042046.1042050}}.

\bibitem[Bra13]{BraultBaron13}
Johann Brault{-}Baron.
\newblock {\em De la pertinence de l'{\'{e}}num{\'{e}}ration : complexit{\'{e}}
  en logiques propositionnelle et du premier ordre. (The relevance of the list:
  propositional logic and complexity of the first order)}.
\newblock PhD thesis, University of Caen Normandy, France, 2013.

\bibitem[B{\"u}c60]{buchi1960weak}
J~Richard B{\"u}chi.
\newblock Weak second-order arithmetic and finite automata.
\newblock {\em Mathematical Logic Quarterly}, 6(1-6), 1960.

\bibitem[CI24]{CapelliI24}
Florent Capelli and Oliver Irwin.
\newblock Direct access for conjunctive queries with negations.
\newblock In Graham Cormode and Michael Shekelyan, editors, {\em 27th
  International Conference on Database Theory, {ICDT} 2024, Paestum, Italy,
  March 25-28, 2024}, volume 290 of {\em LIPIcs}, pages 13:1--13:20. Schloss
  Dagstuhl - Leibniz-Zentrum f{\"{u}}r Informatik, 2024.
\newblock URL: \url{https://doi.org/10.4230/LIPIcs.ICDT.2024.13}, \href
  {https://doi.org/10.4230/LIPICS.ICDT.2024.13}
  {\path{doi:10.4230/LIPICS.ICDT.2024.13}}.

\bibitem[CTG{\etalchar{+}}23]{CarmeliTGKR23}
Nofar Carmeli, Nikolaos Tziavelis, Wolfgang Gatterbauer, Benny Kimelfeld, and
  Mirek Riedewald.
\newblock Tractable orders for direct access to ranked answers of conjunctive
  queries.
\newblock {\em {ACM} Trans. Database Syst.}, 48(1):1:1--1:45, 2023.
\newblock \href {https://doi.org/10.1145/3578517} {\path{doi:10.1145/3578517}}.

\bibitem[CZB{\etalchar{+}}22]{CarmeliZBCKS22}
Nofar Carmeli, Shai Zeevi, Christoph Berkholz, Alessio Conte, Benny Kimelfeld,
  and Nicole Schweikardt.
\newblock Answering (unions of) conjunctive queries using random access and
  random-order enumeration.
\newblock {\em {ACM} Trans. Database Syst.}, 47(3):9:1--9:49, 2022.
\newblock \href {https://doi.org/10.1145/3531055} {\path{doi:10.1145/3531055}}.

\bibitem[DHK22]{DeepHK22}
Shaleen Deep, Xiao Hu, and Paraschos Koutris.
\newblock Ranked enumeration of join queries with projections.
\newblock {\em Proc. {VLDB} Endow.}, 15(5):1024--1037, 2022.
\newblock URL: \url{https://www.vldb.org/pvldb/vol15/p1024-deep.pdf}, \href
  {https://doi.org/10.14778/3510397.3510401}
  {\path{doi:10.14778/3510397.3510401}}.

\bibitem[DKMP22]{DoleschalKMP22}
Johannes Doleschal, Benny Kimelfeld, Wim Martens, and Liat Peterfreund.
\newblock Weight annotation in information extraction.
\newblock {\em Log. Methods Comput. Sci.}, 18(1), 2022.
\newblock URL: \url{https://doi.org/10.46298/lmcs-18(1:21)2022}, \href
  {https://doi.org/10.46298/LMCS-18(1:21)2022}
  {\path{doi:10.46298/LMCS-18(1:21)2022}}.

\bibitem[DSST86]{driscoll1986making}
James~R. Driscoll, Neil Sarnak, Daniel~Dominic Sleator, and Robert~Endre
  Tarjan.
\newblock Making data structures persistent.
\newblock In Juris Hartmanis, editor, {\em Proceedings of the 18th Annual {ACM}
  Symposium on Theory of Computing, May 28-30, 1986, Berkeley, California,
  {USA}}, pages 109--121. {ACM}, 1986.
\newblock \href {https://doi.org/10.1145/12130.12142}
  {\path{doi:10.1145/12130.12142}}.

\bibitem[ECK24]{EldarCK24}
Idan Eldar, Nofar Carmeli, and Benny Kimelfeld.
\newblock Direct access for answers to conjunctive queries with aggregation.
\newblock In Graham Cormode and Michael Shekelyan, editors, {\em 27th
  International Conference on Database Theory, {ICDT} 2024, Paestum, Italy,
  March 25-28, 2024}, volume 290 of {\em LIPIcs}, pages 4:1--4:20. Schloss
  Dagstuhl - Leibniz-Zentrum f{\"{u}}r Informatik, 2024.
\newblock URL: \url{https://doi.org/10.4230/LIPIcs.ICDT.2024.4}, \href
  {https://doi.org/10.4230/LIPICS.ICDT.2024.4}
  {\path{doi:10.4230/LIPICS.ICDT.2024.4}}.

\bibitem[FKRV15]{fagin2015document}
Ronald Fagin, Benny Kimelfeld, Frederick Reiss, and Stijn Vansummeren.
\newblock Document spanners: {A} formal approach to information extraction.
\newblock {\em J. {ACM}}, 62(2):12:1--12:51, 2015.
\newblock \href {https://doi.org/10.1145/2699442} {\path{doi:10.1145/2699442}}.

\bibitem[FMS97]{FrandsenMS97}
Gudmund~Skovbjerg Frandsen, Peter~Bro Miltersen, and Sven Skyum.
\newblock Dynamic word problems.
\newblock {\em J. {ACM}}, 44(2):257--271, 1997.
\newblock \href {https://doi.org/10.1145/256303.256309}
  {\path{doi:10.1145/256303.256309}}.

\bibitem[FRU{\etalchar{+}}20]{florenzano2020efficient}
Fernando Florenzano, Cristian Riveros, Mart{\'{\i}}n Ugarte, Stijn Vansummeren,
  and Domagoj Vrgoc.
\newblock Efficient enumeration algorithms for regular document spanners.
\newblock {\em {ACM} Trans. Database Syst.}, 45(1):3:1--3:42, 2020.
\newblock \href {https://doi.org/10.1145/3351451} {\path{doi:10.1145/3351451}}.

\bibitem[GJ22]{GrandjeanJ22}
Etienne Grandjean and Louis Jachiet.
\newblock Which arithmetic operations can be performed in constant time in the
  {RAM} model with addition?
\newblock {\em CoRR}, abs/2206.13851, 2022.
\newblock URL: \url{https://doi.org/10.48550/arXiv.2206.13851}, \href
  {https://arxiv.org/abs/2206.13851} {\path{arXiv:2206.13851}}, \href
  {https://doi.org/10.48550/ARXIV.2206.13851}
  {\path{doi:10.48550/ARXIV.2206.13851}}.

\bibitem[KMN25]{KleestMeissnerMN22}
Sarah Kleest{-}Mei{\ss}ner, Jonas Marasus, and Matthias Niewerth.
\newblock {MSO} queries on trees: Enumerating answers under updates using
  forest algebras.
\newblock {\em Log. Methods Comput. Sci.}, 21(4), 2025.
\newblock URL: \url{https://doi.org/10.46298/lmcs-21(4:2)2025}, \href
  {https://doi.org/10.46298/LMCS-21(4:2)2025}
  {\path{doi:10.46298/LMCS-21(4:2)2025}}.

\bibitem[Knu98]{Knuth98a}
Donald~Ervin Knuth.
\newblock {\em The art of computer programming, Volume III, 2nd Edition}.
\newblock Addison-Wesley, 1998.

\bibitem[Lib04]{libkin2004elements}
Leonid Libkin.
\newblock {\em Elements of finite model theory}, volume~41.
\newblock Springer, 2004.

\bibitem[LM14]{LosemannM14}
Katja Losemann and Wim Martens.
\newblock {MSO} queries on trees: enumerating answers under updates.
\newblock In Thomas~A. Henzinger and Dale Miller, editors, {\em Joint Meeting
  of the Twenty-Third {EACSL} Annual Conference on Computer Science Logic
  {(CSL)} and the Twenty-Ninth Annual {ACM/IEEE} Symposium on Logic in Computer
  Science (LICS), {CSL-LICS} 2014, Vienna, Austria, July 14 - 18, 2014}, pages
  67:1--67:10. {ACM}, 2014.
\newblock \href {https://doi.org/10.1145/2603088.2603137}
  {\path{doi:10.1145/2603088.2603137}}.

\bibitem[MR24]{MunozR22}
Martin Mu{\~{n}}oz and Cristian Riveros.
\newblock Streaming enumeration on nested documents.
\newblock {\em {ACM} Trans. Database Syst.}, 49(4):15:1--15:39, 2024.
\newblock \href {https://doi.org/10.1145/3701557} {\path{doi:10.1145/3701557}}.

\bibitem[MRV18]{maturana2018document}
Francisco Maturana, Cristian Riveros, and Domagoj Vrgoc.
\newblock Document spanners for extracting incomplete information:
  Expressiveness and complexity.
\newblock In Jan~Van den Bussche and Marcelo Arenas, editors, {\em Proceedings
  of the 37th {ACM} {SIGMOD-SIGACT-SIGAI} Symposium on Principles of Database
  Systems, Houston, TX, USA, June 10-15, 2018}, pages 125--136. {ACM}, 2018.
\newblock \href {https://doi.org/10.1145/3196959.3196968}
  {\path{doi:10.1145/3196959.3196968}}.

\bibitem[Nie18]{Niewerth18}
Matthias Niewerth.
\newblock {MSO} queries on trees: Enumerating answers under updates using
  forest algebras.
\newblock In Anuj Dawar and Erich Gr{\"{a}}del, editors, {\em Proceedings of
  the 33rd Annual {ACM/IEEE} Symposium on Logic in Computer Science, {LICS}
  2018, Oxford, UK, July 09-12, 2018}, pages 769--778. {ACM}, 2018.
\newblock \href {https://doi.org/10.1145/3209108.3209144}
  {\path{doi:10.1145/3209108.3209144}}.

\bibitem[NS18]{NiewerthS18}
Matthias Niewerth and Luc Segoufin.
\newblock Enumeration of {MSO} queries on strings with constant delay and
  logarithmic updates.
\newblock In Jan~Van den Bussche and Marcelo Arenas, editors, {\em Proceedings
  of the 37th {ACM} {SIGMOD-SIGACT-SIGAI} Symposium on Principles of Database
  Systems, Houston, TX, USA, June 10-15, 2018}, pages 179--191. {ACM}, 2018.
\newblock \href {https://doi.org/10.1145/3196959.3196961}
  {\path{doi:10.1145/3196959.3196961}}.

\bibitem[Pet21]{Peterfreund21}
Liat Peterfreund.
\newblock Grammars for document spanners.
\newblock In Ke~Yi and Zhewei Wei, editors, {\em 24th International Conference
  on Database Theory, {ICDT} 2021, Nicosia, Cyprus, March 23-26, 2021}, volume
  186 of {\em LIPIcs}, pages 7:1--7:18. Schloss Dagstuhl - Leibniz-Zentrum
  f{\"{u}}r Informatik, 2021.
\newblock URL: \url{https://doi.org/10.4230/LIPIcs.ICDT.2021.7}, \href
  {https://doi.org/10.4230/LIPICS.ICDT.2021.7}
  {\path{doi:10.4230/LIPICS.ICDT.2021.7}}.

\bibitem[SS21]{SchmidS21}
Markus~L. Schmid and Nicole Schweikardt.
\newblock Spanner evaluation over slp-compressed documents.
\newblock In Leonid Libkin, Reinhard Pichler, and Paolo Guagliardo, editors,
  {\em PODS'21: Proceedings of the 40th {ACM} {SIGMOD-SIGACT-SIGAI} Symposium
  on Principles of Database Systems, Virtual Event, China, June 20-25, 2021},
  pages 153--165. {ACM}, 2021.
\newblock \href {https://doi.org/10.1145/3452021.3458325}
  {\path{doi:10.1145/3452021.3458325}}.

\bibitem[SS22]{SchmidS22}
Markus~L. Schmid and Nicole Schweikardt.
\newblock Query evaluation over slp-represented document databases with complex
  document editing.
\newblock In Leonid Libkin and Pablo Barcel{\'{o}}, editors, {\em {PODS} '22:
  International Conference on Management of Data, Philadelphia, PA, USA, June
  12 - 17, 2022}, pages 79--89. {ACM}, 2022.
\newblock \href {https://doi.org/10.1145/3517804.3524158}
  {\path{doi:10.1145/3517804.3524158}}.

\bibitem[TGR20]{TziavelisGR20}
Nikolaos Tziavelis, Wolfgang Gatterbauer, and Mirek Riedewald.
\newblock Optimal join algorithms meet top-k.
\newblock In David Maier, Rachel Pottinger, AnHai Doan, Wang{-}Chiew Tan,
  Abdussalam Alawini, and Hung~Q. Ngo, editors, {\em Proceedings of the 2020
  International Conference on Management of Data, {SIGMOD} Conference 2020,
  online conference [Portland, OR, USA], June 14-19, 2020}, pages 2659--2665.
  {ACM}, 2020.
\newblock \href {https://doi.org/10.1145/3318464.3383132}
  {\path{doi:10.1145/3318464.3383132}}.

\bibitem[TGR26]{TziavelisGR22}
Nikolaos Tziavelis, Wolfgang Gatterbauer, and Mirek Riedewald.
\newblock Any-k algorithms for enumerating ranked answers to conjunctive
  queries.
\newblock {\em {ACM} Trans. Database Syst.}, 51(1):1:1--1:47, 2026.
\newblock \href {https://doi.org/10.1145/3734517} {\path{doi:10.1145/3734517}}.

\end{thebibliography}

\end{document}